\documentstyle[12pt]{article}
\begin{document}
\title{\LARGE {\bf On the structure of Verma modules over
 Virasoro and Neveu-Schwarz algebras.}}
\author{A. Astashkevich
\thanks{ Supported by Rosenbaum Fellowship}
\thanks{ Department of Mathematics, MIT, Cambridge, MA 02139
\newline
\quad \quad   e-mail address: ast@math.mit.edu}}
\date{October 28, 1995}
\maketitle
\font\got=eufm10 at 12pt
\font\gotb=eufb10 at 12pt
\renewcommand{\floatpagefraction}{2}
\renewcommand{\textfraction}{0}
\def\mod{\mbox{mod}{\,}}
\def\hat{\widehat}
\def\bar{\overline}
\def\Vi{{\cal V}{\it ir}}
\def\Vir{{$\cal V$}{\it ir}}
\def\NV{{\cal NV}}
\def\NeV{{$\cal NV$}}
\section{Introduction}

 The main goal of this paper is to present a different proof of the theorem
of B.~L. Feigin and D.~B. Fuchs (see {\bf [Fe-Fu 1]}) about
 the structure of Verma
modules over Virasoro algebra and to state some
 new results about the structure
of Verma modules over Neveu-Schwarz algebra. This proof has two advantages:
first, it is simpler in the most interesting cases
(for example in the so called
minimal models), and second, it can be generalized
for Neveu-Schwarz algebra for
some class of  Verma modules.

 This text arose as a result of trying to understand the original proof of
B.~L. Feigin and D.~B. Fuchs. The original proof
uses some facts about Jantzen's
filtration which I could not prove and nobody could explain to me.
 That is why I tried to find another proof.

I would like to express my deep gratitude to M. Finkelberg, E. Frenkel and W.
Soergel for valuable discussions. I am happy to thank Professor V.G. Kac for
his interest in my work and his questions.
I am greatly indebted to  D.B. Fuchs for numerous conversations and his
 constant care.

\section{Notation}
\subsection {\bf Virasoro algebra and Verma modules.}
 Let ${\cal L}$ be the Lie algebra of algebraic vector fields on $\bf C^*$ with
 the basis $L_i,~ i{\in}{\bf Z}$ and commutators

 $$\quad\quad\quad  [L_i,L_j]=(j-i)L_{i+j}.$$

 The Virasoro algebra, \Vir , is a one dimensional central extension of
 ${\cal L}$ corresponding to the cocycle $(L_i,L_j){\rightarrow}
{\delta_{-i,j}}{(j^3-j)\over 12}.$ We have the following basis in the Virasoro
algebra $L_i,~ i{\in}{\bf Z}$ and  $C$ and commutators

$$ \quad\quad\quad [L_i,C]=0,$$

$$ \quad\quad\quad [L_i,L_j]=(j-i)L_{i+j} +{\delta_{-i,j}}{(j^3-j)\over 12}C.$$

Both algebras are $\bf Z$-graded: $\deg L_i=i$ and $\deg C=0.$
Let us denote by {\gotb h} a Lie algebra with the basis $L_0$ and $C$,
  by {\gotb n$^-$}
a Lie algebra with the basis \{$L_{-i},~ i{\in}{\bf N}$\},
and by {\gotb n$^+$} a Lie algebra
with the basis \{$L_i,~ i{\in}{\bf N}$\}. We also denote
by {\gotb b$^+$} a Lie algebra with
the basis  $L_i$ and $C,~i{\in}{\bf Z_+}$. All these algebras  {\gotb h},
 {\gotb n$^-$}, {\gotb n$^+$} and  {\gotb b$^+$} are subalgebras of {\Vir}.
We have a Cartan type decomposition of {\Vir}

$$\quad\quad\quad {\Vi}=
{{\mbox{\gotb n}}^-}{\oplus}{\mbox{\gotb h}}{\oplus}{{\mbox{\gotb n}}^+}$$
$$\quad\quad\quad {\Vi}= {{\mbox{\gotb n}}^-}{\oplus}{{\mbox{\gotb b}}^+}.$$

Let $h,c{\in}\bf C$. Let us consider a one
dimensional module $\bf C{_{\em h,c}}$
over    {\gotb b$^+$} such that  {\gotb n$^+$} acts by zero, $L_0$ is
a multiplication by $h$ and $C$ is a multiplication by $c$.
Verma module $M_{h,c}$ over the Virasoro
algebra is by definition an induced module from $\bf C{_{\em h,c}}$

  $${\quad\quad\quad M_{h,c}=
{\mit Ind}^{\Vi}_{{\mbox{\gotb b}}^+}}\bf C_{\em h,c}.$$

We have a natural inclusion of
${\bf C_{\em h,c}}{\hookrightarrow}{M_{h,c}}$. So we have
a vector ${\bf v}{\in}M_{h,c}$ corresponding to
${\bf 1}{\in}{\bf C_{\em h,c}}$. Sometimes we will write ${\bf v_{\em h,c}}$
to stress that this vector lies in ${M_{h,c}}$.  Vector {\bf v} is called
the vacuum vector.

Let us make a few remarks about Verma modules.
First, any Verma module ${M_{h,c}}$ is a free module
over {\bf U}({\gotb n$^-$}). Therefore,
we have the following basis in ${M_{h,c}}$:

  $${L_{-i_k}}{L_{-i_{k-1}}}...{L_{-i_2}}{L_{-i_1}}{\bf v_{\em h,c}}$$
where ${i_k}{\geq}{i_{k-1}}{\geq}...{\geq}{i_1}{\geq}1$.

The operator ${L_0}$ on $M_{h,c}$ is semisimple. We can consider
the eigenspace decomposition of $M_{h,c}$,

$${M_{h,c}=}{\bigoplus_{i=0}^{\infty}}~{M_{h,c}^i},$$
where ${L_0}$ acts as a multiplication by ${h-i}$ on ${M_{h,c}^i}$.
It is easy to see that this decomposition respects the
grading on \Vir. We say that vector ${\bf w}{\in}{M_{h,c}}$
has level {\em n} if ${\bf w}{\in}{M_{h,c}^n}$.

Vector {\bf w} is called singular if it has some
level {\em n} (${\em n{\in}{\bf Z_+}}$) and {\gotb n$^+$} acts
by zero on this vector. It is obvious that any singular vector
generates a submodule isomorphic to Verma module. If a singular
vector has level {\em n} then it generates $M_{h-n,c}$.

\subsection {\bf  Categories ${\cal O}_c,~{\cal O},$ Shapovalov's
form and Kac determinant formula.}

Let us define categories ${\cal O}_c$ and ${\cal O}$.

We say that module ${\bf M}{\in}{\cal O}_c$ if it satisfies the
following conditions:

1) $C$ acts on {\bf M} as a multiplication by {\em c}.

2) $L_0$ acts semisimply on {\bf M} and we have a decomposition of {\bf M}

$${\bf M}={\bigoplus_{h{\in}{\bf C}}}{\bf M}_h,
 ~~~~~where~L_0 ~acts~on ~{\bf M}_h~ as~a~
multiplication~ by~h$$
and $\dim({\bf M}_h)<{\infty}$ for any $h{\in}{\bf C}$.

3) {\gotb n}$^+$ acts locally finite on {\bf M}. This means that for any
${\bf w}{\in}{\bf M} ,~~{\bf U}({\mbox{\gotb n}}^+){\bf w}$ is a
finite dimensional space.

We define category ${\cal O}$ as a direct sum of
${\cal O}_c$ over all $c{\in}{\bf C}$.

{\bf Example:} Module ${M_{h,c}{\in}{{\cal O}_c}}$.

Let us define for every module ${\bf M}{\in}{{\cal O}_c}$ a
contragradient module
${\bar {\bf M}}{\in}{{\cal O}_c}$ in the following way:

$$ {{\bar {\bf M}}_h}{\stackrel {def}=}{{\bf M}^{'}_h}$$
$L_i$ {\it acts on} ${\bar {\bf M}}$ {\it as} ${L^{'}_{-i}}$
{\it and} $C$ {\it acts as a multiplication by} $c$.

It is obvious that
${\bf w}{\stackrel {def}=}{\bf v_{\em h,c}^{'}}{\in}{\bar M}_{h,c}$
 is a singular vector thus we get a  map
${\bf B}:\, {M_{h,c}}{\rightarrow}{M_{h,c}^{'}}$.
This map defines a bilinear form on the module $M_{h,c}$ (Shapovalov's form).
One can show that this form is
  symmetric. By definition this form is contravariant.
It is easy to check that the spaces,
$M_{h,c}^n$ and $M_{h,c}^m$ for $n{\neq}m$ are orthogonal. Since we have a
basis in
$M_{h,c}^n$ we can calculate the determinant of the form ${\bf B}_n$-
restriction of {\bf B} on
$M_{h,c}^n$. The result is well known.

{\bf Kac determinant formula:}

$$  {\det}^2({\bf B_n})={\mit Const}
{\prod_{\begin{array}{c}{\em k,l}\\{\em kl}{\leq}{\bf n}\end{array}}}
{\bf \Phi}_{k,l}(h,c)^{p(n-kl)},~~~~~~~~{\mit where}$$

$$ {\bf \Phi}_{k,l}(h,c)=\left(h+{(k^2-1)(c-13)\over 24}+{(kl-1)\over 2}\right)
\left(h+{(l^2-1)(c-13)\over 24}+{(kl-1)\over 2}\right)$$

$$+{{(k^2-l^2)}^2\over 16}.$$

This formula gives us a condition under which the module $M_{h,c}$ is
irreducible. It is obvious that
if Verma module is reducible then it contains at least one singular vector.

{\bf Remark.} We can define the form {\bf B} in another way.
Let ${\omega}: {\bf U}({\Vi}){\rightarrow}{\bf U}({\Vi})$ be an anti-involution
such that

$$  {\omega}(L_i){\stackrel {def}=}L_{-i}$$
$$   {\omega}(C){\stackrel {def}=}C.$$

Let for ${\bf w}{\in}{M_{h,c}},~~\langle {\bf w}\rangle ~~{\mit be~~the
{}~~vacuum ~expectation~ value}$
$$\langle {\bf w}\rangle {\stackrel {def}=}{\bf a},~~~~~~~~~~{\mit where}$$
$$ {\bf w}={\bf a}{{\bf
v}_{h,c}}+{\sum_{{i_k}{\geq}{i_{k-1}}{\geq}...{\geq}{i_1}{\geq}1}}{\bf
a}_{{i_1},...,{i_k}}{L_{-i_k}}{L_{-i_{k-1}}}...{L_{-i_2}}{L_{-i_1}}{\bf v_{\em
h,c}}.$$

Then for $x,y{\in}{\bf U}({\Vi})$ we have
$$ {\bf B}(x({\bf v_{\em h,c}}),y({\bf v_{\em h,c}})){\stackrel {def}=}
{\langle {\omega}(x)y({\bf v_{\em h,c}})\rangle}.$$

\subsection {\bf References.}  All these facts are  well-known and can be found
in {\bf [Kac-Ra]} or {\bf [Fe-Fu 1]}.

\section{Singular vectors in Verma modules over Virasoro}

{\bf Theorem 3.1. ([Fu 1])}
\begin{it}
  At each level {\em n} only one singular vector {\bf w} can exist.
If a singular vector {\bf w} exists then we have the following formula for it
$${\bf w}={({L_{-1}})^n}{\bf v_{\em h,c}}+
{\sum_{\begin{array}{c}{i_k}+...+{i_1}=n\\{i_k}
{\geq}{i_{k-1}}{\geq}...{\geq}{i_1}{\geq}1\\{i_k{\geq}2}\end{array}}}
{{\bf P}^{(n)}_{{i_1},...,{i_k}}}(h,c)
{L_{-i_k}}{L_{-i_{k-1}}}...{L_{-i_2}}{L_{-i_1}}{\bf v_{\em h,c}}$$
which defines {\bf w} up to  multiplication by a constant.
${{\bf P}^{(n)}_{{i_1},...,{i_k}}}(h,c)$ are polynomials in  h  and  c.
\end{it}

{\bf Proof:}
   First of all, any element ${\bf w}{\in}{M_{h,c}^n}$
can be written in a unique way in the form:
 $${\bf w}={\bf a}{({L_{-1}})^n}{\bf v_{\em h,c}}+
{\sum_{\begin{array}{c}{i_k}+...+{i_1}=n\\{i_k}
{\geq}{i_{k-1}}{\geq}...{\geq}{i_1}{\geq}1\\{i_k{\geq}2}\end{array}}}
{\bf a}_{{i_1},...,{i_k}}{L_{-i_k}}{L_{-i_{k-1}}}...{L_{-i_2}}{L_{-i_1}}
{\bf v_{\em h,c}}.$$

    Let us order the monomials,
${L_{-i_k}}{L_{-i_{k-1}}}...{L_{-i_2}}
{L_{-i_1}}$ (${i_k}{\geq}{i_{k-1}}{\geq}...{\geq}{i_1}{\geq}1$),
in the following way.
 We say that ${L_{-i_k}}{L_{-i_{k-1}}}...{L_{-i_2}}{L_{-i_1}}$ is greater
 then ${L_{-j_l}}{L_{-j_{l-1}}}...{L_{-j_2}}{L_{-j_1}}$ if
${i_1}={j_1}, {i_2}={j_2},...,{i_m}={j_m}$ and
${i_{m+1}}>{j_{m+1}}$ for some ${\em m}$.

{\bf Example:} ${({L_{-1}})^n}<{L_{-2}}{({L_{-1}})^{n-2}}<
{L_{-3}}{({L_{-1}})^{n-3}}<.\,.\,. $

 Let us assume that {\bf w} is a  singular vector.
This means that for any ${i{\geq}1}$
${L_{i}}{\bf w}=0$. Our goal is to express all
coefficients ${\bf a}_{{i_1},...,{i_k}}$
as

 ${\bf a}{\times}{\mit (some~polynomial~depending~only ~on ~h~and~c)}$.

We will show that this can be done by induction. If we know all coefficients
${\bf a}_{{i_1},...,{i_k}}$ for
${L_{-i_k}}{L_{-i_{k-1}}}...{L_{-i_2}}{L_{-i_1}}<{L_{-j_l}}
{L_{-j_{l-1}}}...{L_{-j_2}}{L_{-j_1}}$ then we can express the
coefficient ${\bf a}_{{j_1},...,{j_l}}$
as a sum of ${\bf a}_{{i_1},...,{i_k}}{{\bf Q}_{{i_1},...,{i_k}}(h,c)}$,
 where ${{\bf Q}_{{i_1},...,{i_k}}(h,c)}$ are polynomials in $h$
and $c$.

Let ${j_1}={j_2}=...={j_s}=1$ and ${j_{s+1}}>1$. Then let us calculate the
coefficient
$$\mbox{at} ~~~{L_{-j_l}}{L_{-j_{l-1}}}...{L_{-j_{s+2}}}
{(L_{-1})^{s+1}}{\bf v_{\em h,c}}~~~\mbox{in}~~~{L_{j_{s+1}-1}}{\bf w}.$$

One can see that it has the following form:
$$(2{j_{s+1}}-1){\bf a}_{{j_1},...,{j_l}}+$$
$$+{\sum_{\begin{array}{c}{i_k}{\geq}{i_{k-1}}{\geq}...{\geq}{i_1}
{\geq}1\\{L_{-i_k}}{L_{-i_{k-1}}}...{L_{-i_2}}{L_{-i_1}}<{L_{-j_l}}
{L_{-j_{l-1}}}...{L_{-j_2}}{L_{-j_1}}\end{array}}}{\bf a}_{{i_1},...,{i_k}}
{{\bf Q}_{{i_1},...,{i_k}}(h,c)},$$
where ${\bf a}_{1,...,1}={\bf a}$. Since this coefficient equals zero we can
express  ${\bf a}_{{j_1},...,{j_l}}$  as a sum of  ${\bf a}_{{i_1},...,{i_k}}
{{\bf Q}_{{i_1},...,{i_k}}(h,c)}$,  where
${{\bf Q}_{{i_1},...,{i_k}}(h,c)}$  are polynomials in $h$ and $c$,  and
the sum is over such indices  ${i_1},...,{i_k}$  that
$${L_{-i_k}}{L_{-i_{k-1}}}...{L_{-i_2}}{L_{-i_1}}<{L_{-j_l}}
{L_{-j_{l-1}}}...{L_{-j_2}}{L_{-j_1}}.$$
By induction we immediately get that  ${\bf a}_{{j_1},...,{j_l}}$
is of the form
 $${\bf a}{\times}{\mit (some~polynomial~depending~only ~on ~h,c)}.$$
 From this the proposition follows immediately.

 \quad \quad \quad \quad \quad \quad \quad \quad \quad \quad \quad \quad \quad
\quad \quad \quad \quad \quad \quad \quad \quad \quad        {\bf  Q.E.D.}

  Now let us look at the Kac determinant formula:
$$  {\det}^2({\bf B_n})={\mit Const}
{\prod_{\begin{array}{c}
{\em k,l}\\
{\em kl}{\leq}{\bf n}
\end{array}}}
{\bf \Phi}_{ k,l}(h,c).$$

Equation ${\bf \Phi}_{\em k,l}(h,c)=0$ defines a rational curve $({\bf C}^*)$
in ${\bf C}^2$.
 This curve can be given by the formulas:
$$h(t)={{1-k^2}\over 4}t+{{1-kl}\over 2}+{{1-l^2}\over 4}t^{-1}{\,},$$
$$c(t)=6t+13+6t^{-1}.$$
Denote this curve by ${\cal F}(k,l)$.

The Kac determinant formula shows that for almost all $t{\in}{\bf C}^*$
(except finite number)
there exist a singular vector in ${M_{h(t),c(t)}}$ at level {\em kl}.
We get the following corollary of proposition 3.1.

{\bf Corollary 3.2. ([Fu 1])}
\begin {it}
 For any natural numbers k and l there exist a unique map
${\bf S}_{k,l}:{\cal F}(k,l){\rightarrow}{{\bf U}({\mbox {\gotb n}}^-)}_{kl}$
which has the following form:
$${\bf
%% FOLLOWING LINE CANNOT BE BROKEN BEFORE 80 CHAR
S}_{k,l}(t)={\sum_{\begin{array}{c}{i_k}+...+{i_1}=kl\\{i_r}{\geq}{i_{r-1}}{\geq}...{\geq}{i_1}{\geq}1\end{array}}}{{\bf P}^{k,l}_{{i_1},...,{i_r}}}(t){L_{-i_r}}{L_{-i_{r-1}}}...{L_{-i_2}}{L_{-i_1}}{\bf v_{\em h,c}}$$
where ${\bf P}_{1,...,1}^{k,l}{\equiv}1$ and
${\bf P}^{k,l}_{{i_1},...,{i_r}}(t)$ is a
polynomial in $t$ and $t^{-1}$ for any ${{i_1},...,{i_r}}$
 and such that ${\bf S}_{k,l}(t){\bf v_{\em h,c}}$ is a singular vector in the
module
${M_{h(t),c(t)}}$, where h(t) and c(t) are given by the formulas above
({\it i.e.}, t  is a
parameter on the curve ${\cal F}(k,l)$).
\end {it}

{\bf Proof:}
  Follows immediately from the proposition 3.1.

 \quad \quad \quad \quad \quad \quad \quad \quad \quad \quad \quad \quad \quad
\quad \quad \quad \quad \quad \quad \quad \quad \quad        {\bf  Q.E.D.}

We have a trivial vector bundle ${{\bf U}({\mbox {\gotb n}}^-)}_{kl}$
over ${\bf C}P^1$ and we have a section ${\bf S}_{k,l}(t)$ of this bundle
 over ${\bf C}^*$. Consider this section as a meromorfic section of
our vector bundle over ${\bf C}P^1$. Now we would like to calculate the orders
of the poles at points zero and infinity.

Let us formulate the final result. The proof of the following theorem is
technical and can be found in {\bf [Ast-Fu]}.

{\bf Theorem 3.3.}
\begin {it}
  The coefficient at ${L_{-i_r}}{L_{-i_{r-1}}}...{L_{-i_2}}{L_{-i_1}}$ in
${\bf S}_{k,l}(t)$
has degree in $t$ less or equal than $l(k-1)$.
The degree in $t$ is equal to $l(k-1)$ only at the
monomial  $(L_k)^l$ and the coefficient at $t^{l(k-1)}$ equals $((k-1)!)^l$.
\end {it}

 As a corollary of the last theorem one get the following important result.

{\bf Theorem 3.4.}
\begin {it}
  The orders of the poles of the section ${\bf S}_{k,l}(t)$ of the trivial
vector bundle ${{\bf U}({\mbox {\gotb n}}^-)}_{kl}$ over ${\bf C}P^1$ are equal
to $l(k-1)$ at $\infty$ and $k(l-1)$ at 0.
\end {it}

{\bf Proof:}  Obvious.

 \quad \quad \quad \quad \quad \quad \quad \quad \quad \quad \quad \quad \quad
\quad \quad \quad \quad \quad \quad \quad \quad \quad        {\bf  Q.E.D.}

\section{Jantzen's filtration.}

 The main goal of this section is to define Jantzen's filtration and to
formulate all properties of it
which we need.

 Let $\cal C$ be a smooth algebraic curve over {\bf C}
with the sheaf of functions $\cal O$ and two vector bundles
  $M$ and $\bar M$.
 Denote the corresponding sheaves by  ${\cal M}$ and $\bar {\cal M}$.
Suppose we have a map ${\bf B}:\,M{\rightarrow}{\bar M}$ of
the vector bundles .
 Then for any point $p{\in}{\cal C}$ we  get  Jantzen's
filtration on the fiber
of {\em M} at point $p$. Let us define it. Let $\tau$ be a local
parameter at point $p{\in}{\cal C}$.
Consider the ring ${\cal O}_p$ and the modules
${\cal M}_p$ and ${\bar {\cal M}}_p$
over it. These modules are free and we have a map
${\bf B}_p:{{\cal M}}_p{\rightarrow}{\bar {\cal M}}_p$.
 The fibers of  $M$ and $\bar M$ at point $p$ are exactly
 ${\cal M}_p/{\tau}{\cal M}_p$ and
${\bar {\cal M}}_p/{\tau}{\bar {\cal M}}_p$.
We denote them by $V$ and $W$ respectively.
Now we will define a decreasing
filtration $\cdots V^{(2)}{\subseteq}V^{(1)}{\subseteq}V^{(0)}=V$.
\newline

{\bf Definition:}
\begin {it}
  $ V^{(n)}$ is spanned by such vectors $v$  that
there exists an element ${\tilde v}{\in}{\cal M}_p$
with the following properties:

{\bf i)} Under the projection
${\cal M}_p{\stackrel {\pi}{\rightarrow}}{\cal M}_p/{\tau}{\cal M}_p=V$
$~~\pi({\tilde v})=v$.

{\bf ii)} ${\bf B}_p({\tilde v}){\in}{\tau}^n{\bar {\cal M}}_p$.
\end {it}

{}From the definition it is obvious that the filtration depends only on the
map {\bf B} in some neighborhood
of point $p$.

Suppose we have a symmetric bilinear form {\bf B} on the
vector bundle $M$.
Take $\bar M=M^{'}$- the dual vector bundle. Bilinear form provides a map
$M{\rightarrow}M^{'}={\bar M}$. So we obtain Jantzen's filtration on every
fiber
of the vector bundle $M$.
{}From now on
we assume that we
have a vector bundle $M$ and a bilinear form {\bf B} on it.
Choose a point $p{\in}{\cal C}$ and let us denote the
induced map from $V{\rightarrow}V^{'}$
by ${\bf B}_{(p)}$ where $V$ is a fiber of $M$ at point $p$.
\linebreak

{\bf Properties of Jantzen's filtration.}

 {\bf 1)} $V^{(1)}=Ker({\bf B}_{(p)})$

 {\bf 2)} Let us assume that we have two maps $A$ and
$A^{'}:{M}{\rightarrow}{M}$
such that ${\bf B}(Av,w)={\bf B}(v,A^{'}w)$ for
any two sections $v,w{\in}{\Gamma}({\cal U},{\cal M})$
where ${\cal U}$ is any open  subset of ${\cal C}$.
We have induced maps $A_{(p)}~ ~and~~A^{'}_{(p)}:V{\rightarrow}V$ and
it is easy to check that
$$A_{(p)}(V^{(n)}){\subseteq}V^{(n)}.$$

Inded, if $v{\in}V^{(n)}$ then there exists ${\tilde v}{\in}{\cal M}_p$
such that
${\bf B}_p({\tilde v}){\in}{\tau}^n{\cal M}^{'}_p$.
This means that for any $w{\in}{\cal M}_p~~$
${\bf B}_p({\tilde v},w){\in}{\tau}^n{\cal O}_p$.
Therefore, in order to check that $A_{(p)}v{\in}V^{(n)}$
it is sufficient to show that  for any $w{\in}{\cal M}_p$
${\bf B}_p({A_p}{\tilde v},w){\in}{\tau}^n{\cal O}_p$.  But
${\bf B}_p({A_p}{\tilde v},w)={\bf B}_p({\tilde
v},{A^{'}_p}w){\in}{\tau}^n{\cal O}_p$ $~$
because  ${A^{'}_p}w{\in}{\cal M}_p$ for any $w{\in}{\cal M}_p$ .

  {\bf 3)} Let us assume that the form {\bf B} is non-degenerate at
the generic point. Then we
can define a determinant of this form as a section of the following line bundle
$({\Lambda}^{\dim~M}M)^{\otimes 2}$.
$$\det({\bf B}){\in}{\Gamma}({\cal C},
(({\Lambda}^{\dim~M}M)^{\otimes 2})').$$
We have the following formula:
$$ord_{\tau}(\det({\bf B}))={\sum_{i=1}^{\infty}}~\dim(V^{(i)}).$$

 The statement is local, so to prove it, it is enough to consider a
free module ${\cal M}_p$ over
${\cal O}_p$ and a bilinear form ${\bf B}_p$. In such case the formula is
almost obvious.
One can find a proof of it in the Jantzen's book (see {\bf [Ja]}).

 {\bf 4)} Assume that the form {\bf B} is symmetric and non-degenerate
at generic
 point. Then it induces non-degenerate
symmetric bilinear form on each quotient
 $V^{(i)}/V^{(i+1)}$ where $i{\in}{\bf N}.$

 The statement is local. One can find the proof of it in Jantzen's book (see
{\bf [Ja]}).

  {\bf 5)}  Assume that we have a smooth algebraic surface, $\cal S$,
over {\bf C}, a
point, $p{\in}{\cal S}$, and two smooth curves ${\cal C}_1$ and
${\cal C}_2$ which
 intersect transversally at this point.
 Suppose we have a vector bundle $M$ over $\cal S$ and a bilinear
form {\bf B}  on $M$ which is non-degenerate at
the generic  point of the surface
${\cal S}$. We will denote the fiber of the vector bundle $M$ at point $p$
by $V$. In addition, assume that we have two sections
${\tilde v}$ and  ${\tilde w}{\in}{\Gamma}({\cal S},M)$ such that

 {\bf a)} $v={\tilde v}(p)={\tilde w}(p){\in}V$

 {\bf b)} ${\bf B}|_{{\cal C}_1}({\tilde v}|_{{\cal C}_1},\bullet)=0$  and
          ${\bf B}|_{{\cal C}_2}({\tilde w}|_{{\cal C}_2},\bullet)=0$.

   Let ${\cal C}$ be any smooth curve which contains point $p$.
If we restrict the vector
bundle $M$ and the form {\bf B} to $\cal C$
we obtain Jantzen's filtration
on the vector space $V$ if we restrict our vector
bundle $M$ and the form {\bf B} to $\cal C$.
The claim is
$$v{\in}V^{(2)}.$$

  Proof:  The statement is local,
 therefore we can consider a free module ${\cal M}_p$ over the local ring
 ${\cal O}_{{\cal S},p}$. Moreover,
we can take completions with respect to  the
maximal ideal {\gotb m}${\subset}{\cal O}_{{\cal S},p}$.
 Then ${\hat {\cal O}}_{{\cal S},p}{\cong}{\bf C}[[x,y]]$
and we can assume that the curves ${\cal C}_1$ and $ {\cal C}_2$ are given by
the equations
$x=0$ and $y=0$ correspondingly. The equation for the curve
$\cal C$ is $z=0$ where
$z=ax+by+{\tilde z}$ and ${\tilde z}{\in}{\mbox {\gotb m}}^2$.
We have form $\hat {{\bf B}_p}$ on the module $\hat {M_p}$.

 Now we forget about the curve and reformulate everything
in purely algebraic terms. We have a ring ${\bf C}[[x,y]]$, a
free module ${\hat V}=V{\otimes}{\bf C}[[x,y]]$, and a map
${\hat {\bf B}}:{\hat V}
{\rightarrow}{\hat V}^{'}$ over ${\bf C}[[x,y]]$. Moreover,
we have two vectors,
${\tilde v},{\tilde w}{\in}{\hat V}$, such that modulo the
maximal ideal {\gotb m}
$\subset {\bf C}[[x,y]]$ they are equal to
$v{\in}V={\hat V}/{\mbox {\gotb m}}{\hat V}$ and
${\bf B}(\tilde v){\in}(x){\hat V}^{'},
{}~~{\bf B}(\tilde w){\in}(y){\hat V}^{'}.$
We can write down the map {\bf B} as a Taylor series:
 $${\bf B}={\bf B}_0+x{\bf B}_{1,0}+y{\bf B}_{0,1}+O({\mbox {\gotb m}}^2).$$
Also we can write
 $$\tilde v=v+yv_{0,1}+xv_{1,0}+O({\mbox {\gotb m}}^2)~~~~~\mbox{and}$$
$$\tilde w=v+yw_{0,1}+xw_{1,0}+O({\mbox {\gotb m}}^2).$$
 Then we have the following equalities:
 $${\bf B}_0(v)=0,~~{\bf B}_{0,1}(v)={\bf B}_0(v_{0,1}),~~{\bf B}_{1,0}(v)
={\bf B}_0(w_{1,0}).$$
 We must show that we can find $u{\in}{\hat V}$ such that
$u=v~~~\mod{\,}{\mbox {\gotb m}}$ and
${\bf B}(u)=0~~~\mod{\,}(z=ax+by+\tilde z,{\mbox {\gotb m}}^2)=
(ax+by,{\mbox {\gotb m}}^2)$.
 But this is obvious since we can solve the following equation:
 $$x{\bf B}_{1,0}(v)+y{\bf B}_{0,1}(v)
=-{\bf B}_0(xu_{1,0}+yu_{0,1})~~~\mod{\,}(ax+by,{\mbox {\gotb m}}^2).$$

 \quad \quad \quad \quad \quad \quad \quad \quad \quad \quad \quad \quad
\quad  \quad \quad \quad \quad \quad \quad \quad \quad \quad
      {\bf  Q.E.D.}

\section {Structure  of  submodules  of Verma  modules and
Jantzen's filtration.}

  In this section we will state the main theorems.

  Let us fix $h$ and $c$. Then equation ${\Phi}_{k,l}(h,c)=0$ defines in plane
${\bf C}^2(k,l)$
 a quadruple of straight lines namely $pk+ql+m=0$,  where
$$c={{(3p+2q)(3q+2p)}\over{pq}},~~~~~~h={{-m^2-(p+q)^2}\over{4pq}}.$$
Certainly $p$, $q$ and $m$ are not defined
uniquely but nevertheless, the lines $pk+ql+m=0$ in plane
${\bf C}^2(k,l)$ are correctly defined.
It is obvious that the directions of these lines depend only
 on $c$. When $c{\neq}1,25$, these lines form
a rhombus with the diagonals $k=\pm l$. If $c=1$ (or $c=25$) then this
rhombus degenerates into a pair of lines parallel
to the line $k=l$ (or $k=-l$) and symmetric with
respect to this  line. Moreover, if
$c=1$ and $h=0$  (or $c=25$ and $h=-1$ ) then these
two lines become a single line $k=l$
 (or $k=-l$ ). These lines are real if and only
if $c{\leq}1$ or $c{\geq}25$. If $c{\leq}1$
 then all these lines have positive slope and when
$c{\geq}25$ all these lines have negative
slope. These lines are never parallel to
 coordinate axis. Let us denote one of this lines
by $l_{h,c}$.

\subsection{\bf  Structure of submodules of Verma modules.}

We shall distinguish the following cases.

\begin{itemize}
\item[{\bf Case I.}]   The line $l_{h,c}$ contains no integral points.

\item[{\bf Case II.}]  The line $l_{h,c}$ contains exactly one integral
point $(k,l)$. We have following subcases:

\begin{itemize}
\item[${\bf II_{+}.}$]  The product $kl>0$.

\item[${\bf II_{0}.}$]  The product $kl=0$.

\item[${\bf II_{-}.}$]  The product $kl<0$.
\end{itemize}

\item[{\bf Case III.}]  The line $l_{h,c}$ contains infinitely many integral
points. We will distinguish six subcases which can be divided in two groups.

\item[${\bf Subcase ~~c{\leq}1.}$]
 Let $(k_1,l_1),(k_2,l_2),(k_3,l_3),...$
be all integral points $~(k,l)$ on the line $~l_{h,c}$
up to equivalence $~(k,l){\sim}(k^{'},l^{'})~$ iff
$kl=k^{'}l^{'}$ and such that $kl>0$.
We ordered them in such a way that ${k_i}{l_i}<{k_{i+1}}{l_{i+1}}$ for
all $i{\in}{\bf N}$.

\begin{itemize}
\item[${\bf III_{-}^{00}.}$] Line $l_{h,c}$ intersects
both axes at integral points
 (see Figure 1).

\item[${\bf III_{-}^{0}.}$]  Line $l_{h,c}$ intersects
only one axis at integral point
(see Figure 2).

\item[${\bf III_{-}.}$]   Line $l_{h,c}$ intersects both
axes at non-integral points.
In this case we draw an auxiliary line $l_{h,c}^{'}$ parallel
to $l_{h,c}$ through the point
$(k_1,-l_1)$. We denote its points
$(k_2,{l_2}-2{l_1}),(k_3,{l_3}-2{l_1}),(k_4,{l_4}-2{l_1}),...$
by  $(k_1^{'},l_1^{'}),(k_2^{'},l_2^{'})$, $(k_3^{'},l_3^{'}),...~$
 (see Figure 3).
It is easy to see that we have  the following inequalities:
$${k_1}{l_1}<{k_2}{l_2}<{k_1}{l_1}+{k_1^{'}}{l_1^{'}}<{k_1}{l_1}+{k_2^{'}}
{l_2^{'}}<{k_3}{l_3}<{k_4}{l_4}<$$
$$<{k_1}{l_1}+{k_3^{'}}{l_3^{'}}<{k_1}{l_1}+{k_4^{'}}
{l_4^{'}}<{k_5}{l_5}<{k_6}{l_6}<...~~ .$$
\end{itemize}

\item[ ${\bf Subcase~~ c{\geq}25.}$]
 Let ${\{}(k_1,l_1),(k_2,l_2),...,(k_s,l_s){\}}$
 be all integral points $(k,l)$
on the line $l_{h,c}$ up to equivalence
$(k,l){\sim}(k^{'},l^{'})~~\mbox{iff}~~kl=k^{'}l^{'}$ and such that $kl>0$.
We ordered them in such a way that ${k_i}{l_i}<{k_{i+1}}{l_{i+1}
}$ for all $i{\in}{\{}1,2,...,s-1{\}}$.

\begin{itemize}
\item[${\bf III_{+}^{00}.}$] Line $l_{h,c}$ intersects
 both axes at integral points
(see Figure 4).

\item[${\bf III_{+}^{0}.}$]  Line $l_{h,c}$ intersects
 only one axis at integral point
(see Figure 5).

\item[${\bf III_{+}.}$]   Line $l_{h,c}$ intersects
 both axes at non-integral points.
In this case we draw an auxiliary line $l_{h,c}^{'}$ parallel
to $l_{h,c}$ through the point
$(k_1,-l_1)$. We denote its points
$(k_2,{l_2}-2{l_1})$, $(k_3,{l_3}-2{l_1})$,
$(k_4,{l_4}-2{l_1}),{\,}.{\,}.{\,}.$
by
$(k_1^{'},l_1^{'})$, $(k_2^{'},l_2^{'})$,
$(k_3^{'},l_3^{'}),{\,}.{\,}.{\,}.~$
(see Figure 6).
 It is easy to see that we have  the following inequalities:
$${k_1}{l_1}<{k_2}{l_2}<{k_1}{l_1}+{k_1^{'}}{l_1^{'}}
<{k_1}{l_1}+{k_2^{'}}{l_2^{'}}<{k_3}{l_3}<{k_4}{l_4}<$$
$$<{k_1}{l_1}+{k_3^{'}}{l_3^{'}}<{k_1}{l_1}+
{k_4^{'}}{l_4^{'}}<{k_5}{l_5}<{k_6}{l_6}<...~~ .$$
\end{itemize}
\end{itemize}

{\bf Theorem A.  ([Fe-Fu 1])}
\begin {it}

{\bf a)} All submodules of Verma module are generated by singular vectors.

{\bf b)} \quad i) In  cases $\bf I,~II_{-}$ and $\bf II_{0}$ Verma module
is irreducible.

\quad \quad ii) In  case $\bf II_{+}$ Verma module $M_{h,c}$ has
a unique submodule generated
by the singular vector at level ${\,}kl$. This submodule
is isomorphic to Verma module $M_{h-kl,c}$ which
is irreducible (case $\bf II_{-}$).

\quad \quad iii) In  cases $\bf III^{00}_{-},~III^{0}_{-}$ and $\bf III_{-}$
we have an infinite number
 of singular vectors. All singular vectors and relations
between them are shown in the diagrams below. Singular vectors are
denoted by points with
 their weights indicated. An arrow or a chain of arrows from one
point to another
means that the second singular vector vector lies in the submodule
generated by the first
one.

\quad \quad i{\rm v})  In
cases $\bf III^{00}_{+},~III^{0}_{+}$ and $\bf III_{+}$
 we have a finite number
 of singular vectors (maybe zero). All singular vectors and relations
between them are shown in the diagrams below. Singular vectors are denoted
by points with
 their weights indicated. An arrow or a chain of arrows from one
point to another
means that the second singular vector  lies in the submodule
generated by the first
one.

{\bf c)} In  case
$\bf III^{00}_{-}~~($respectively~$\bf III^{0}_{-},~III_{-},
{}~III^{00}_{+},~III^{0}_{+}$ and $\bf III_{+})$
any Verma submodule generated by a singular vector belongs to  case
$\bf ~~III^{00}_{-}$ {\hfill}
{\bf (}respectively $\bf III^{0}_{-},~III_{-},~III^{00}_{+},
{}~III^{0}_{+}$ and $\bf III_{+}).$

\end {it}

\begin{figure}[t!]
\setlength{\unitlength}{0.012500in}%
\begin{picture}(355,285)( -30,530)
\thicklines
\put( 60,665){\circle*{10}}
\put( 60,710){\circle*{10}}
\put( 60,755){\circle*{10}}
\put(235,785){\circle*{10}}
\put(200,750){\circle*{10}}
\put(200,710){\circle*{10}}
\put(200,670){\circle*{10}}
\put(200,630){\circle*{10}}
\put(200,590){\circle*{10}}
\put(270,750){\circle*{10}}
\put(270,710){\circle*{10}}
\put(270,670){\circle*{10}}
\put(270,630){\circle*{10}}
\put(270,590){\circle*{10}}
\put(265,625){\vector(-2,-1){ 60}}
\put( 60,750){\vector( 0,-1){ 35}}
\put( 60,705){\vector( 0,-1){ 35}}
\put( 60,660){\vector( 0,-1){ 35}}
\put(200,705){\vector( 0,-1){ 30}}
\put(270,705){\vector( 0,-1){ 30}}
\put(265,705){\vector(-2,-1){ 60}}
\put(205,705){\vector( 2,-1){ 60}}
\put(200,665){\vector( 0,-1){ 30}}
\put(270,665){\vector( 0,-1){ 30}}
\put(200,745){\line( 0,-1){ 25}}
\put(200,720){\vector( 0,-1){  5}}
\put(270,745){\vector( 0,-1){ 30}}
\put(265,745){\vector(-2,-1){ 60}}
\put(205,745){\vector( 2,-1){ 60}}
\put(240,780){\vector( 1,-1){ 25}}
\put(230,780){\vector(-1,-1){ 25}}
\put(205,665){\vector( 2,-1){ 60}}
\put(265,665){\vector(-2,-1){ 60}}
\put(200,625){\vector( 0,-1){ 30}}
\put(270,625){\line( 0,-1){ 25}}
\put(270,600){\line( 0, 1){  0}}
\put(270,600){\vector( 0,-1){  5}}
\put(205,625){\vector( 2,-1){ 60}}
\put(265,625){\vector(-2,-1){ 60}}
\put(200,585){\line( 0,-1){ 10}}
\put(270,585){\line( 0,-1){ 10}}
\put( 20,790){\makebox(0,0)[lb]{\smash{\Large {$\bf and~~~  III^{\mbox
 {\small {\bf 0}}}_{\bf -}$}}}}
\put( 50,610){\makebox(0,0)[lb]{\smash{\Large {\bf .}}}}
\put( 60,610){\makebox(0,0)[lb]{\smash{\Large {\bf .}}}}
\put( 70,610){\makebox(0,0)[lb]{\smash{\Large {\bf .}}}}
\put( 15,810){\makebox(0,0)[lb]{\smash{\Large {$\bf Cases~~~III^{\mbox
 {\small {\bf 00}}}_{\bf -}
{}~~~~~~~~~~Case~~~~III_{\bf -}$}}}}
\put(210,570){\makebox(0,0)[lb]{\smash{\Large {\bf   .   .   .}}}}
\put(185,785){\makebox(0,0)[lb]{\smash{\small {\bf (h , c)}}}}
\put( 15,755){\makebox(0,0)[lb]{\smash{\small {\bf (h , c)}}}}
\put(85,625){\makebox(0,0)[lb]{\smash{\small {$\bf
(h-k_{1}l_{1}-k^{'}_{3}l^{'}_{3},c)$}}}}
\put(85,705){\makebox(0,0)[lb]{\smash{\small {$\bf
(h-k_{1}l_{1}-k^{'}_{1}l^{'}_{1},c)$}}}}
\put(280,710){\makebox(0,0)[lb]{\smash{\small {$\bf
(h-k_{1}l_{1}-k^{'}_{2}l^{'}_{2},c)$}}}}
\put(280,630){\makebox(0,0)[lb]{\smash{\small {$\bf
(h-k_{1}l_{1}-k^{'}_{4}l^{'}_{4},c)$}}}}
\put(110,665){\makebox(0,0)[lb]{\smash{\small {$\bf (h-k_{3}l_{3},c)$}}}}
\put(110,750){\makebox(0,0)[lb]{\smash{\small {$\bf (h-k_{1}l_{1},c)$}}}}
\put(110,590){\makebox(0,0)[lb]{\smash{\small {$\bf (h-k_{5}l_{5},c)$}}}}
\put(285,750){\makebox(0,0)[lb]{\smash{\small {$\bf (h-k_{2}l_{2},c)$}}}}
\put(285,670){\makebox(0,0)[lb]{\smash{\small {$\bf (h-k_{4}l_{4},c)$}}}}
\put(285,590){\makebox(0,0)[lb]{\smash{\small {$\bf (h-k_{6}l_{6},c)$}}}}
\put(-15,710){\makebox(0,0)[lb]{\smash{\small {$\bf (h-k_{1}l_{1},c)$}}}}
\put(-15,660){\makebox(0,0)[lb]{\smash{\small {$\bf (h-k_{2}l_{2},c)$}}}}
\end{picture}
\end{figure}
\begin{figure}[t!]
\setlength{\unitlength}{0.012500in}
\begin{picture}(290,247)(-30,600)
\thicklines
\put( 70,775){\circle*{10}}
\put( 70,745){\circle*{10}}
\put( 70,715){\circle*{10}}
\put( 70,605){\circle*{10}}
\put( 70,635){\circle*{10}}
\put(230,790){\circle*{10}}
\put(205,760){\circle*{10}}
\put(255,760){\circle*{10}}
\put(205,730){\circle*{10}}
\put(255,730){\circle*{10}}
\put(205,700){\circle*{10}}
\put(255,700){\circle*{10}}
\put(205,670){\circle*{10}}
\put(255,670){\circle*{10}}
\put(220,650){\circle*{2}}
\put(230,650){\circle*{2}}
\put(240,650){\circle*{2}}
\put( 60,680){\circle*{2}}
\put( 60,680){\circle*{2}}
\put( 60,680){\circle*{2}}
\put( 70,680){\circle*{2}}
\put( 80,680){\circle*{2}}
\put(205,615){\circle*{10}}
\put(255,615){\circle*{10}}
\put(230,585){\circle*{10}}
\put( 70,770){\vector( 0,-1){ 20}}
\put( 70,740){\vector( 0,-1){ 20}}
\put( 70,630){\vector( 0,-1){ 20}}
\put( 70,710){\line( 0,-1){ 10}}
\put( 70,660){\vector( 0,-1){ 20}}
\put(225,785){\vector(-1,-1){ 20}}
\put(235,785){\vector( 1,-1){ 20}}
\put(205,755){\vector( 0,-1){ 20}}
\put(255,755){\vector( 0,-1){ 20}}
\put(210,755){\vector( 2,-1){ 40}}
\put(250,755){\vector(-2,-1){ 40}}
\put(205,725){\vector( 0,-1){ 20}}
\put(255,725){\vector( 0,-1){ 20}}
\put(250,725){\vector(-2,-1){ 40}}
\put(250,755){\vector(-2,-1){ 40}}
\put(250,755){\vector(-2,-1){ 40}}
\put(210,725){\vector( 2,-1){ 40}}
\put(250,695){\vector(-2,-1){ 40}}
\put(210,695){\vector( 2,-1){ 40}}
\put(205,695){\vector( 0,-1){ 20}}
\put(255,695){\vector( 0,-1){ 20}}
\put(205,665){\line( 0,-1){ 10}}
\put(255,665){\line( 0,-1){ 10}}
\put(205,690){\line( 0, 1){  0}}
\put(205,690){\line( 0, 1){  0}}
\put(205,640){\vector( 0,-1){ 20}}
\put(230,640){\vector(-1,-1){ 20}}
\put(255,640){\vector( 0,-1){ 20}}
\put(230,640){\vector( 1,-1){ 20}}
\put(250,610){\vector(-1,-1){ 20}}
\put(210,610){\vector( 1,-1){ 20}}
\put( 20,825){\makebox(0,0)[lb]{\smash{\Large {$\bf Cases~~~III^{\mbox {\small
{\bf 00}}}_{\bf +}$}}}}
\put( 20,795){\makebox(0,0)[lb]{\smash{\Large {$\bf and~~~III^{\mbox {\small
{\bf 0}}}_{\bf +}$}}}}
\put(200,825){\makebox(0,0)[lb]{\smash{\Large {$\bf Case ~~~III_{\bf +}$}}}}
\put( 15,770){\makebox(0,0)[lb]{\smash{\small {$\bf (h,c)$}}}}
\put( -5,740){\makebox(0,0)[lb]{\smash{\small {$\bf (h-k_{1}l_{1},c)$}}}}
\put( -5,715){\makebox(0,0)[lb]{\smash{\small {$\bf (h-k_{2}l_{2},c)$}}}}
\put(-30,630){\makebox(0,0)[lb]{\smash{\small {$\bf (h-k_{s-1}l_{s-1},c)$}}}}
\put( -5,600){\makebox(0,0)[lb]{\smash{\small {$\bf (h-k_{s}l_{s},c)$}}}}
\put(130,755){\makebox(0,0)[lb]{\smash{\small {$\bf (h-k_{1}l_{1},c)$}}}}
\put( 90,725){\makebox(0,0)[lb]{\smash{\small {$\bf
(h-k_{1}l_{1}-k^{'}_{1}l^{'}_{1},c)$}}}}
\put(130,695){\makebox(0,0)[lb]{\smash{\small {$\bf (h-k_{3}l_{3},c)$}}}}
\put( 90,660){\makebox(0,0)[lb]{\smash{\small {$\bf
(h-k_{1}l_{1}-k^{'}_{3}l^{'}_{3},c)$}}}}
\put(165,785){\makebox(0,0)[lb]{\smash{\small {$\bf (h,c)$}}}}
\put(265,755){\makebox(0,0)[lb]{\smash{\small {$\bf (h-k_{2}l_{2},c)$}}}}
\put(265,725){\makebox(0,0)[lb]{\smash{\small {$\bf
(h-k_{1}l_{1}-k^{'}_{2}l^{'}_{2},c)$}}}}
\put(265,695){\makebox(0,0)[lb]{\smash{\small {$\bf (h-k_{4}l_{4},c)$}}}}
\put(265,665){\makebox(0,0)[lb]{\smash{\small {$\bf
(h-k_{1}l_{1}-k^{'}_{4}l^{'}_{4},c)$}}}}
\end{picture}
\end{figure}

\subsection{\bf Jantzen's filtration.}

Firsts of all, let us make  some remarks about the
curves ${\cal F}(k,l)$. Curves
${\cal F}(k,l)$ intersect each other only at real points.
  Curves ${\cal F}(k,k)$
are lines $ h={{k^2-1}\over{24}}(1-c) $.
All other curves ${\cal F}(k,l)$ for $k{\neq}l$
have two branches in  real plane ${\bf R^{2}}$.
One of the branches of the curve ${\cal F}(k,l)$ lies in
the region $c{\leq}h{\geq}{{(c-1)}\over{24}}$ the other one lies in the region
$c{\geq}25~~h{\leq}0$. All curves ${\cal F}(k,l)$ for $k{\neq}l$
touch boundary lines
$c=1,~~h={{(c-1)}\over{24}}$ and $c=25$.

We have a two parameter family of Verma modules $M_{h,c}$.
Moreover, we have a symmetric
bilinear form  ${\bf B}(h,c)$ on our Verma module $M_{h,c}$. We
can restrict  the
 form ${\bf B}(h,c)$ to $M_{h,c}^i$ where $i{\in}{\bf Z_{+}}$.
We denote this restriction by  ${\bf B}_i(h,c)$.
One can look at this situation in
the following way. We have a complex algebraic surface
${\bf C^2}$ and a trivial vector
bundle ${\bf M}_{i}$ with fibers which are canonically
isomorphic to $M_{h,c}^{i}$ at point
$(h,c){\in}{\bf C^2}$.
We have a symmetric bilinear form on the vector bundle
${\bf M}_{i}$ which is non-degenerate at the generic point of ${\bf C^2}$.
The Kac
determinant formula gives us expression for the determinant
of this form in some basis.

Let us fix a point $(h,c){\in}{\bf C^2}$.
 Let ${\cal C}$ be any smooth curve which passes
through this point $(h,c)$. Consider Jantzen's filtration on the fiber of the
vector bundle ${\bf M}_{i}$ at point $(h,c)$ {\it i.e.}, $M_{h,c}^i$
along this curve
 ${\cal C}$. From the properties of
Janten'z filtration (see section 4.) follow
that the filtration
is \Vir\ invariant. Let us describe it. We assume here that the
curve ${\cal C}$
is not tangent to any curve ${\cal F}(k,l)$ at point $(h,c)$.
We will keep the same notation as in Theorem A.

{\bf Theorem B.}
\begin {it}

{\bf a)} In  cases ${\bf I,~II_{\mbox {\small {\bf 0}}}}$ and $\bf II_{-}$ all
 $M_{h,c}^{(i)}=0~~$ for $i{\in}{\bf N}$.

{\bf b)} In  case ${\bf II_{+}}$ we have one point $(k,l)$ on
the line $l_{h,c}$ such that
$kl>0$.
 $M_{h,c}^{(1)}$ is generated by the singular vector at  level $~~kl~$
and is isomorphic to Verma module $M_{h-kl,c}$.
$M_{h,c}^{(i)}=0~~for~~i>1,~~i{\in}{\bf N}$.

{\bf c)} In  case ${\bf III_{-}^{\mbox {\small {\bf 0}}}}$ we have
an infinite number of
points $(k_i,l_i)$ on the line $l_{h,c}$.
 Submodule $M_{h,c}^{(i)}$ is generated by the singular vector at
level $~~k_il_i~~$ and is isomorphic to Verma module $M_{h-k_il_i,c}$.

{\bf d)} In case ${\bf III_{+}^{\mbox {\small {\bf 0}}}}$ we have
a finite number of
points $(k_1,l_1), ... ,(k_s,l_s)$ on the line $l_{h,c}$.
Submodule $M_{h,c}^{(i)}$ is generated by the singular vector at
level $~~k_il_i~~$ and is isomorphic to Verma module
$M_{h-k_il_i,c}$ for $i{\leq}s$.
$M_{h,c}^{(i)}=0~~$ for $i>s$.

{\bf e)} In  case ${\bf III_{-}}$ we have an infinite number of
points $(k_i,l_i)$ on the line $l_{h,c}$
and points $(k_j^{'},l_j^{'})$ on the additional line.
Submodule $M_{h,c}^{(2i-1)}$ is generated by two singular vectors
at levels $k_{2i-1}l_{2i-1}~~and$ $k_{2i}l_{2i}$ for $i{\in}{\bf N}$.
 Submodule $M_{h,c}^{(2i)}$ is generated by two singular vectors at levels
$k_1l_1+k_{2i-1}^{'}l_{2i-1}^{'}~~and$ $k_1l_1+k_{2i}^{'}l_{2i}^{'}$
for $i{\in}{\bf N}$.

{\bf f)} In  case ${\bf III_{+}}$ we have a finite number of
points $(k_1,l_1), ... ,(k_s,l_s)$ on the line $l_{h,c}$
and points $(k_1^{'},l_1^{'}), ... ,(k_{s-1}^{'},l_{s-1}^{'})$
on the additional
line. Submodule $M_{h,c}^{(2i-1)}$ is
 generated by two singular vectors at levels $k_{2i-1}l_{2i-1}$
and $k_{2i}l_{2i}$ for
$2i{\leq}s$.  Submodule $M_{h,c}^{(2i)}$ is
generated by two singular vectors at levels
$k_1l_1+k_{2i-1}^{'}l_{2i-1}^{'}~~and$ $k_1l_1+k_{2i}^{'}l_{2i}^{'}$  for
$2i<s$.

 We have two case (which depend on s being even or odd):

 {\quad} {\quad} s is even)  Submodule $M_{h,c}^{(s)}$ is generated by
the singular vector at  level  $k_1l_1+k_{s-1}^{'}l_{s-1}^{'}$
and is isomorphic to Verma module
$M_{h-k_1l_1+k_{s-1}^{'}l_{s-1}^{'},c}$.
For all $i>s$ submodule $M_{h,c}^{(i)}=0$.

 {\quad} {\quad} s is odd) Submodule $M_{h,c}^{(s)}$ is generated by the
singular vector at  level $k_sl_s$
and is isomorphic to Verma module $M_{h-k_sl_s,c}$.
For all $i>s$ submodule $M_{h,c}^{(i)}=0$.

{\bf g)} In  case  ${\bf III_{-}^{\mbox {\small {\bf 00}}}}$ we have
an infinite number of
points $(k_i,l_i)$ on the line $l_{h,c}$.
We distinguish two cases:

 {\quad} {\quad} $c=1~~or~~c=24h+1$ ) Submodule $M_{h,c}^{(i)}$ is generated
by the singular vector at  level $k_il_i$
and is isomorphic to Verma module $M_{h-k_il_i,c}$.

{\quad} {\quad} $c{\neq}1~~and~~c{\neq}24h+1$ ) Submodule
 $M_{h,c}^{(2i-1)}=M_{h,c}^{(2i)}$ is generated
by the singular vector at level $k_il_i$
and is isomorphic to Verma module $M_{h-k_il_i,c}$.

{\bf h)} In  case  ${\bf III_{+}^{\mbox {\small {\bf 00}}}}$  we have
a finite number of
points $(k_1,l_1), ... ,(k_s,l_s)$ on the line $l_{h,c}$.
We distinguish two cases:

 {\quad} {\quad} $c=25$ ) Submodule $M_{h,c}^{(i)}$ is generated
by the singular vector at  level $k_il_i$
and is isomorphic to Verma module $M_{h-k_il_i,c}$
for $1{\leq}i{\leq}s$. For all $i>s$ we have  $M_{h,c}^{(i)}=0$.

 {\quad} {\quad} $c{\neq}25$ )  Submodule
 $M_{h,c}^{(2i-1)}=M_{h,c}^{(2i)}$ is generated
by the singular vector at  level $k_il_i$
and is isomorphic to Verma module $M_{h-k_il_i,c}$
for $1{\leq}i{\leq}s$.  For all $i>2s$ we have  $M_{h,c}^{(i)}=0$.

\end {it}

 Now we will describe Jantzen's filtration in the case when
 ${\cal C}={\cal F}(\bar k,\bar l)$.
We have a fixed point
$(h,c){\in}{\bf C^2}$ and we assume that $(h,c){\in}{\cal C}={\cal F}(\bar
k,\bar l)$. We will keep the same notation as in Theorem A.

{\bf Theorem C.}
\begin {it}

{\bf a)} Cases ${\bf I,~II_{\mbox {\small {\bf 0}}}}$ and $\bf II_{-}$
cannot occur.

{\bf b)} In  case ${\bf II_{+}}$ we have a point $(k,l)$
on the line $l_{h,c}$ such that
$kl>0$ and $(\bar k,\bar l)=(k,l)$. Then $M_{h,c}^{(i)}=M_{h,c}^{(1)}$
 is generated by the singular vector at level $~~kl~$ and is
isomorphic to Verma module $M_{h-kl,c}$ for all $i{\in}{\bf N}$.

{\bf c)} In  case ${\bf III_{-}^{\mbox {\small {\bf 0}}}}$ we have
an infinite number of
points $(k_i,l_i)$ on the line $l_{h,c}$ and for some $j{\in}{\bf N}$
$(k_j,l_j)=(\bar k,\bar l)$.
Submodule $M_{h,c}^{(i)}$ is generated by the singular vector at
level $~~k_il_i~~$ and is isomorphic to Verma
module $M_{h-k_il_i,c}$ for $i{\leq}j$.
For $i>j$  $M_{h,c}^{(i)}=M_{h,c}^{(j)}$.

{\bf d)} In case ${\bf III_{+}^{\mbox {\small {\bf 0}}}}$ we have
a finite number of
points $(k_1,l_1), ... ,(k_s,l_s)$ on the line $l_{h,c}$.
For some $j{\in}{\{}1,...,s{\}}$ $(k_j,l_j)=(\bar k,\bar l)$.
Submodule $M_{h,c}^{(i)}$ is generated by the singular vector at
 level $~~k_il_i~~$ and is isomorphic
to Verma module $M_{h-k_il_i,c}$ for $i{\leq}j$.
$M_{h,c}^{(i)}=M_{h,c}^{(j)}$ for $i>j$.

{\bf e)} Case ${\bf III_{-}}$. There is an infinite number of points
$(k_j,l_j)$ on the line  $l_{h,c}$.
For some $j{\in}{\bf N}$  $(k_j,l_j)=(\bar k,\bar l)$.
Submodule $M_{h,c}^{(2i-1)}$ is generated by two singular vectors
at  levels $k_{2i-1}l_{2i-1}$ and
$k_{2i}l_{2i}$ for $2i-1{\leq}j$. Submodule
$M_{h,c}^{(2i)}$ is generated by two singular vectors at  levels
$k_1l_1+k_{2i-1}^{'}l_{2i-1}^{'}$ and
$k_1l_1+k_{2i}^{'}l_{2i}^{'}$  for $2i<j$.
Let's denote by $\tilde j=\cases{
j+1&if $j$ is odd \cr
j&if $j$ is even \cr}$. Then for any $i{\geq}{\tilde j}$
$M_{h,c}^{(i)}$ is  generated by the singular vector at level $k_jl_j$
and is isomorphic to Verma module $M_{h-k_jl_j,c}$.

{\bf f)} In case ${\bf III_{+}}$ we have a finite number of
points $(k_1,l_1), ... ,(k_s,l_s)$ on the line $l_{h,c}$
and points $(k_1^{'},l_1^{'}), ... ,(k_{s-1}^{'},l_{s-1}^{'})$
on the additional
line. For some $1{\leq}j{\leq}s$  $(k_j,l_j)=(\bar k,\bar l)$.
 Submodule $M_{h,c}^{(2i-1)}$ is
 generated by two singular vectors at
levels $k_{2i-1}l_{2i-1}~~and$ $k_{2i}l_{2i}$ for
$2i-1{\leq}j$.  Submodule $M_{h,c}^{(2i)}$ is
generated by two singular vectors at  levels
$k_1l_1+k_{2i-1}^{'}l_{2i-1}^{'}~~and$ $k_1l_1+k_{2i}^{'}l_{2i}^{'}$  for
$2i<j$. Let's denote by $\tilde j=\cases{
j+1&if $j$ is odd \cr
j&if $j$ is even \cr}$. Then for any $i{\geq}{\tilde j}$
$M_{h,c}^{(i)}$ is generated by the
singular vector at  level $k_jl_j$ and is
 isomorphic to Verma module $M_{h-k_jl_j,c}$.

{\bf g)} In case  ${\bf III_{-}^{\mbox {\small {\bf 00}}}}$
we have an infinite number of
points $(k_i,l_i)$ on the line $l_{h,c}$.
There exists $j{\in}{\bf N}$ such that $k_jl_j={\bar k}{\bar l}$.
Submodule $M_{h,c}^{(2i-1)}=M_{h,c}^{(2i)}$ is generated
by the singular vector at  level $k_il_i$
and is isomorphic to Verma module $M_{h-k_il_i,c}$
for $1{\leq}i{\leq}j$. For any $i>2j$  $M_{h,c}^{(i)}=M_{h,c}^{(2j)}$.

{\bf h)} In  case  ${\bf III_{+}^{\mbox {\small {\bf 00}}}}$
we have a finite number of
points $(k_1,l_1), ... ,(k_s,l_s)$ on the line $l_{h,c}$.
There exists $j$, $1{\leq}j{\leq}s$, such that $k_jl_j={\bar k}{\bar l}$.
Submodule $M_{h,c}^{(2i-1)}=M_{h,c}^{(2i)}$ is generated
by the singular vector at level $k_il_i$
and is isomorphic to Verma module $M_{h-k_il_i,c}$
for $1{\leq}i{\leq}j$. For any $i>2j$  $M_{h,c}^{(i)}=M_{h,c}^{(2j)}$.
\end {it}

\subsection{\bf Remarks.}

  {\bf 1)} We will prove these theorems in two steps.
Step 1.) We  prove sections {\bf b)} (i), (ii) and (iii) of the
 theorem A and the section
{\bf a)} of  theorem A in these cases. Alongside we prove sections
a) through f) of theorem B. Then as a corollary we
 get that parts
a) through f) of theorem C are true.
Step 2.) We  prove sections {\bf b)} (iv) of theorem A and the section
{\bf a)} of theorem A in the above case. Alongside we  prove sections
g) and h) of theorem C. As a corollary we obtain that sections
g) and h) of theorem B are true.

  {\bf 2)} We prove everything by induction by level. Let us explain what
 it means. We say that some property is true up to level $~k~$ if this property
holds in
$${\bigoplus_{i=0}^{k}}~{M_{h,c}^i}.$$
For example, we say that some submodule $V$ is generated
by a vector $v$ up to level
$~k~$ iff
$${\bigoplus_{i=0}^{k}}~{M_{h,c}^i}{\cap}V
={\bigoplus_{i=0}^{k}}~{M_{h,c}^i}{\cap}W,~~~~{\mit where}~
W~{\mit is~a~submodule~generated~by}~~v.$$
Another example. We say that part a) of theorem A holds up to
level $~k~$ meaning that any submodule
of Verma module is generated by singular vectors up to level $~k~$.

   {\bf 3)} It is easy to show (using the Kac determinant formula)
that theorem B follows from
theorem A and vice versa.

\section {Proof of the structure theorem in simple cases.}

In this section we will prove the
structure theorem in the following cases: {\bf I}, {\bf II},
${\bf III_{+}^{\mbox {\small {\bf 0}}}}$,
 ${\bf III_{-}^{\mbox {\small {\bf 0}}}}$,
${\bf III_{+}}$ and ${\bf III_{-}}$. Let us emphasize
that in all these cases  all curves ${\cal F}(k,l)$ which pass through point
$(h,c)$ intersect transversally at this point.
First, let us notice that from the Kac determinant
formula and corollary 3.2  immediately
follows the existence of all the singular vectors
and the diagrams of inclusions between
the corresponding Verma modules
 as stated in the Theorem A. Also, it is easy to see that
any Verma submodule generated by a singular vector (which we constructed)
is of the same case as the original Verma module. Our goal is to show that the
maximal submodule is generated by the
singular vectors (or singular vector) at  levels
 $k_1l_1$ and $k_2l_2$ (or level $k_1l_1$ ) in the notation of theorem A.

Second, let us notice that cases {\bf I}, ${\bf II_{-}}$ and ${\bf II_{0}}$ are
trivial.
They immediately follow from the
Kac determinant formula (since in these cases determinant
does not vanish).

{\bf Definition 6.1.}

\begin {it}
For any module ${\bf M}{\in}{\cal O}_{c}$ let us define its character
$$ch({\bf M}){\stackrel {def}=}{\sum_{h{\in}{\bf C}}}\dim({\bf M}_h)q^h,$$
$${\mit where}~~~{\bf M}={\bigoplus_{h{\in}{\bf C}}}{\bf M}_h.$$
\end {it}

Using property 3 of Jantzen's filtration (section 4)
we can write an explicit
formula for ${\sum_{i=1}^{\infty}}ch(M_{h,c}^{(i)})$. In cases
${\bf III_{-}^{\mbox {\small {\bf 0}}}}$ and ${\bf III_{-}}$
we obtain the following
formula:
$${\sum_{i=1}^{\infty}}ch(M_{h,c}^{(i)})=
{\sum_{i=1}^{\infty}}ch(M_{h-k_il_i,c}).$$
In cases ${\bf III_{+}^{\mbox {\small {\bf 0}}}}$ and
${\bf III_{+}}$ we have only a finite
number of marked points $(k_1,l_1),...,$ $(k_s,l_s)$ on the line $l_{h,c}$.
So we obtain the following formula:
$${\sum_{i=1}^{\infty}}ch(M_{h,c}^{(i)})={\sum_{i=1}^s}ch(M_{h-k_il_i,c}).$$
In case  ${\bf II_{+}}$ we get
$${\sum_{i=1}^{\infty}}ch(M_{h,c}^{(i)})=ch(M_{h-kl,c})$$
where $(k,l)$ is the marked point on the line $l_{h,c}$.

Let us make the following important remark. Jantzen's filtration is a
filtration
by \Vir\ submodules (this follows immediately from the second property of
Jantzen's filtration). Therefore, for example, if we know some vector
$w{\in}M_{h,c}^{(i)}$ then we see that the submodule generated by
the  vector $w$
is contained in $M_{h,c}^{(i)}$.

Let us show that in case  ${\bf II_{+}}$ Verma module $M_{h,c}$
has a unique submodule
$M_{h-kl,c}$. Certainly, we know that such submodule exists
(since we have a singular
vector at level $kl$).
{}From the Kac determinant formula follows that module $M_{h-kl,c}$
is irreducible. Since the kernel of the form is exactly
the maximal submodule,
it contains our submodule $M_{h-kl,c}$.
{}From the property 1 of Jantzen's filtration
follows that $M_{h-kl,c}{\subset}M_{h,c}^{(1)}$. Comparing formula
$${\sum_{i=1}^{\infty}}ch(M_{h,c}^{(i)})=ch(M_{h-kl,c})$$
with the fact that $ch(M_{h,c}^{(1)}){\geq}ch(M_{h-kl,c})$ we obtain
that $M_{h,c}^{(1)}=M_{h-kl,c}$ and $M_{h,c}^{(i)}=0$ for all $i>1$.

Now let us prove theorems A and B for case ${\bf III_{-}}$.
The proof for other
cases is the same with minor modifications. We will use the
same notation as in theorems A and B.
It is useful to look at figures 7 and 8
for a better understanding of the proof.

First of all, let us make the following remarks. Submodule generated by
singular vectors at levels $k_1l_1$ and $k_2l_2$  is
contained in $M_{h,c}^{(1)}$. The more important fact is that
the module generated by the singular vectors at
levels $k_1l_1+k_1^{'}l_1^{'}$
and $k_1l_1+k_2^{'}l_2^{'}$  is contained in $M_{h,c}^{(2)}$.
This fact follows from property 5
of Jantzen's filtration, since each of these two singular vectors comes along
the curves ${\cal F}(k_1,l_1)$ and ${\cal F}(k_2,l_2)$.
This  immediately implies
(comparing these remarks with the formula
for ${\sum_{i=1}^{\infty}}ch(M_{h,c}^{(i)})$)
that the structure of submodules of Verma module $M_{h,c}$ is exactly as
stated in theorems A
and B  up to level $\min(k_3l_3,k_4l_4)-1$.

\begin{figure}[t!]
\setlength{\unitlength}{0.012500in}%
\begin{picture}(250,217)(80,620)
\thicklines
\put(240,625){\circle*{2}}
\put(250,625){\circle*{2}}
\put(260,625){\circle*{2}}
\put(240,800){\line(-1,-2){ 90}}
\put(240,800){\line( 1,-2){ 90}}
\put(235,780){\line(-1,-2){ 80}}
\put(235,780){\line( 1,-2){ 80}}
\put(250,770){\line(-1,-2){ 75}}
\put(250,770){\line( 1,-2){ 75}}
\put(240,740){\line(-1,-2){ 60}}
\put(240,740){\line( 1,-2){ 60}}
\put(250,730){\line(-1,-2){ 55}}
\put(250,730){\line( 1,-2){ 55}}
\put(240,670){\line(-3,-5){ 30}}
\put(240,670){\line( 2,-3){ 33.846}}
\put(260,675){\line( 3,-5){ 30}}
\put(260,675){\line(-3,-5){ 30}}
\multiput(160,710)(8.10811,0.00000){19}{\line( 1, 0){  4.054}}
\put(310,710){\line( 0, 1){  0}}
\put(310,710){\line( 0, 1){  0}}
\multiput(140,690)(8.00000,0.00000){23}{\line( 1, 0){  4.000}}
\put(150,800){\makebox(0,0)[lb]{\smash{$Case~~\alpha )$}}}
\put(280,820){\makebox(0,0)[lb]{\smash{\Large {\bf Figure  7.}}}}
\put( 80,710){\makebox(0,0)[lb]{\smash{level  n}}}
\put( 80,685){\makebox(0,0)[lb]{\smash{level  n+1}}}
\end{picture}
\end{figure}
\begin{figure}[t]
\setlength{\unitlength}{0.0100in}%
\begin{picture}(340,274)(-40,565)
\thicklines
\put(180,805){\line( 2,-3){162.308}}
\put(170,755){\line(-2,-3){128.462}}
\put(170,755){\line( 2,-3){128.462}}
\put(200,740){\line(-2,-3){120}}
\put(200,740){\line( 2,-3){120}}
\put(180,680){\line(-2,-3){ 80}}
\put(180,680){\line( 2,-3){ 80}}
\put(210,660){\line(-3,-4){ 75}}
\put(210,660){\line( 2,-3){ 66.154}}
\put(220,595){\line(-3,-4){ 25.800}}
\put(220,595){\line( 3,-4){ 25.800}}
\multiput( 25,595)(8.02469,0.00000){41}{\line( 1, 0){  4.012}}
\put(180,805){\line(-2,-3){162.308}}
\multiput( 25,610)(7.97468,0.00000){40}{\line( 1, 0){  3.987}}
\put(205,560){\makebox(0,0)[lb]{\smash{\bf ...}}}
\put(190,620){\line(-3,-4){ 45}}
\put(190,620){\line( 2,-3){ 40}}
\multiput( 70,555)(4.90029,7.35043){28}{\makebox(0.4444,0.6667){.}}
\multiput(200,755)(4.95727,-7.43590){28}{\makebox(0.4444,0.6667){.}}
\multiput(335,555)(-9.13793,0.00000){30}{\makebox(0.4444,0.6667){.}}
\multiput(280,560)(-5.00000,7.50000){15}{\makebox(0.4444,0.6667){.}}
\multiput(210,665)(-5.00000,-5.00000){3}{\makebox(0.4444,0.6667){.}}
\multiput(200,655)(-5.00000,7.50000){5}{\makebox(0.4444,0.6667){.}}
\multiput(180,685)(-4.93213,-7.39819){18}{\makebox(0.4444,0.6667){.}}
\multiput( 95,560)(9.00000,0.00000){6}{\makebox(0.4444,0.6667){.}}
\multiput(140,560)(5.46667,7.28889){10}{\makebox(0.4444,0.6667){.}}
\multiput(190,625)(4.87180,-7.30770){10}{\makebox(0.4444,0.6667){.}}
\multiput(235,560)(9.00000,0.00000){6}{\makebox(0.4444,0.6667){.}}
\put(-10,615){\makebox(0,0)[lb]{\smash{level  n}}}
\put(-25,580){\makebox(0,0)[lb]{\smash{level  n+1}}}
\put( 40,785){\makebox(0,0)[lb]{\smash{$Case~~\beta )~~j=1$}}}
\put(230,820){\makebox(0,0)[lb]{\smash{\Large {\bf Figure  8.}}}}
\end{picture}
\end{figure}

We will  prove  theorem A together with
theorem B for this case (for all modules)
 by induction by level.
Certainly, we have the base of induction.
Assume that we proved those theorems up to level $~n$.
Let us prove them up to level $~n+1$ for module $M_{h,c}$.

Let us introduce two additional filtrations ${\bf F}^{(i)}$ and
${\bf G}^{(i)}$. ${\bf F}^{(2i-1)}$ is generated by the singular vectors
at levels $k_{2i-1}l_{2i-1}$ and $k_{2i}l_{2i}$
and ${\bf F}^{(2i)}$ is generated
 by the singular vectors at levels $k_1l_1+k_{2i-1}^{'}l_{2i-1}^{'}$ and
$k_1l_1+k_{2i}^{'}l_{2i}^{'}$. What we want to prove is that the
filtration ${\bf F}^{(i)}$
coincides with the filtration $M_{h,c}^{(i)}$.

${\bf G}^{(i)}$ is generated by the vectors ${\bf F}^{(i)}{\cap}
{\bigoplus_{j=1}^n}M_{h,c}^j$.
It is  important to understand how the filtration
${\bf G}^{(i)}$ looks. In some sense it is a
cutting of the filtration ${\bf F}^{(i)}$.
We know that, up to level $n$, ${\bf G}^{(i)}$ coincides with
$M_{h,c}^{(i)}$. Let us also notice the following properties:
$${\bf G}^{(i)}{\subset}M_{h,c}^{(i)}~~~ \mbox{and}~~~{\bf G}^{(i)}
{\subset}{\bf F}^{(i)}~~~\mbox{for~~any}~~i{\in}{\bf N}.$$
We distinguish two cases.

Case $\alpha$): $n+1{\neq}k_il_i,k_1l_1+k_i^{'}l_i^{'}$
for all $i{\in}{\bf N}$.
Then we  see that the formulas for  ${\sum_{i=1}^{\infty}}ch(M_{h,c}^{(i)})$
and ${\sum_{i=1}^{\infty}}ch({\bf G}^{(i)})$ coincide up to level $~n+1$.
Together with the fact that ${\bf F}^{(i)}={\bf G}^{(i)}$
up to level $~n+1$ in this case
(direct check) we obtain that the filtration  ${\bf F}^{(i)}$
coincides with the filtration $M_{h,c}^{(i)}$ up to level $~n+1$. This proves
the statements of  theorems A and B up to level $~n+1$ for $M_{h,c}$.

Case $\beta$): $n+1=k_il_i~~or~~k_1l_1+k_i^{'}l_i^{'}$
for some $i{\in}{\bf N}$.
Then direct calculations (very easy) show that
$$Res_{q}(q^{-h+n}({\sum_{j=1}^{\infty}}ch(M_{h,c}^{(j)})
-{\sum_{j=1}^{\infty}}ch({\bf G}^{(j)}))dq)=1.$$
Therefore, there exists $j{\in}{\bf N}$ such that
$\dim((M_{h,c}^{(j)})^{n+1}/({\bf G}^{(j)})^{n+1})=1.$
We will use the following notation
$$\tilde i=\cases{
i-1&if $n+1=k_il_i$ and $i$ is odd \cr
i-2&if $n+1=k_il_i$ and $i$ is even \cr
i&if $n+1=k_1l_1+k_i^{'}l_i^{'}$ and $i$ is odd \cr
i-1&if $n+1=k_1l_1+k_i^{'}l_i^{'}$ and $i$ is even \cr}.$$
We distinguish three cases: first $j=1$,
 second $1<j{\leq}{\tilde i}~$ and  third $j={\tilde i}+1$.

The third case is exactly the statement of
theorem B. From the induction hypothesis one
can see that $M_{h,c}^{(j)}$ is exactly as stated in
the theorem (since everything is contained in
$M_{h,c}^{(1)}$ and this ``reduces'' level ).
We know that
$$Res_{q}(q^{-h+n}({\sum_{j=1}^{\infty}}ch(M_{h,c}^{(j)})
-{\sum_{j=1}^{\infty}}ch({\bf G}^{(j)}))dq)=1$$
$$\mbox{and}~~~Res_{q}(q^{-h+s}({\sum_{j=1}^{\infty}}ch(M_{h,c}^{(j)})
-{\sum_{j=1}^{\infty}}ch({\bf G}^{(j)}))dq)=0~~\mbox{for}~~s<n.$$
Therefore, the same argument as in the third case proves that
the second case can not occur.
(since everything is contained in $M_{h,c}^{(1)}$ and this ``reduces'' level ).

Thus, we must show that only the first case can not occur.
If ${\tilde i}=1$ then everything
follows from the remarks at the beginning of the proof.
Thus, we can assume that ${\tilde i}>1$.
Since we assumed that $\dim((M_{h,c}^{(1)})^{n+1}/({\bf G}^{(1)})^{n+1})=1$
 we have
${\bf G}^{(s)}=M_{h,c}^{(s)}$ up to level $n+1$ for all $s>1$. Therefore
$V{\stackrel {def}=}M_{h,c}^{({\tilde i})}/M_{h,c}^{({\tilde i}+1)}$
$={\bf G}^{({\tilde i})}/{\bf G}^{({\tilde i}+1)}$ up to level $~n+1$.
{}From  property 4 of Jantzen's filtration we obtain
a symmetric non-degenerate bilinear form
on module V. We know that (in terms of module $M_{h,c}$ )
module V is generated by two
singular vectors up to level $n+1$.
 So up to level $n+1$ this form is determined
by its values on these two singular vectors.
Let $V_1$ and $V_2$ be two submodules of $V$  generated
 by the first and the second singular vectors respectively.
By the induction hypothesis we know that these
submodules intersect at level $n+1$
and the intersection (at this level) is one dimensional and
is generated by the singular
vector. Since $V_1$ and $V_2$ are quotients of
Verma modules we see that the
form on them must coincide with the
standard form (up to a constant multiple). It is zero on
their intersection (since form is zero on the maximal submodule).
So we see (since $V=V_1+V_2$ up to level $n+1$) that the
singular vector at level $n+1$ lies in the
kernel of the form.
This contradicts  the fact that the form is non-degenerate.
We proved that case $j=1$ is impossible.

Thus we proved the statements of
theorems A and B for case ${\bf III_{-}}$.
Using similar arguments one can prove theorems A and B for cases
${\bf III_{+}^{\mbox {\small {\bf 0}}}}$,
${\bf III_{-}^{\mbox {\small {\bf 0}}}}$
and ${\bf III_{+}}$.

In figures 7 and 8 we draw Verma modules as cones. One cone is contained
 in the other if
the same is true for the corresponding Verma modules. By dotted lines
we marked
Verma module generated by the singular vector at
 level $k_2l_2$  and the quotient
$V{\stackrel {def}=}M_{h,c}^{({\tilde i})}/M_{h,c}^{({\tilde i}+1)}
={\bf G}^{({\tilde i})}/{\bf G}^{({\tilde i}+1)}$ assuming that $j=1$.
One can see
the structure of this  module in the figure, so it becomes clear that such
 situation is impossible.

\section {Completion of the proof.}
\subsection{\bf General remarks and the idea of the proof.}

  In this section we
 will prove theorems A, B and C for the remaining cases. Our
proof is a minor modification of the proof of Feigin and Fuchs
(see {\bf [Fe-Fu 1]}).
First of all, let us make the following remark.
If theorem A is true up to level $n$
then theorems B and C are also true up to level $n$.
This is obvious, since we have an
explicit formula for ${\sum_{j=1}^{\infty}}ch(M_{h,c}^{(j)})$
and we know   the structure
of submodules of Verma module up to level $n$.
This  shows that
there is a unique possibility for Jantzen's filtration.
Therefore, theorems B and C are true
up to level $n$. The next remark is that in order to prove theorem A
up to level $n+1$ (we assume  that
it is true up to level $n$) it is enough to
show that in cases ${\bf III_{+}^{\mbox {\small {\bf 00}}}}$ and
 ${\bf III_{-}^{\mbox {\small {\bf 00}}}}$ the
maximal submodule is generated by
the singular vector at level $k_1l_1$ up to level $n+1$.
Indeed,  all submodules
 are contained in the submodule generated by
the singular vector at the level $k_1l_1$ up to level $n+1$
{\it i.e.}, in Verma module $M_{h-k_1l_1,c}$ up
to level $n+1$ with respect to the original
Verma module $M_{h,c}$
(in other words up to level $~n+1-k_1l_1$ with respect to $M_{h-k_1l_1,c}$).
Therefore, we can apply the induction hypothesis.

We will be proving theorems A, B and C by
induction ``up to level $n$''. Now we assume
that they are true up to level $n$. From
the previous remarks it is enough to show that
in cases ${\bf III_{+}^{\mbox {\small {\bf 00}}}}$ and
 ${\bf III_{-}^{\mbox {\small {\bf 00}}}}$ the
maximal submodule is generated by
the singular vector at  level $k_1l_1$ up to level $~n+1$.
We distinguish two cases.

$\alpha )$ $~n+1{\neq}k_il_i$ for all $i{\in}{\bf N}$ (or $1{\leq}i{\leq}s$).
Let us take some smooth curve through
the point $(h,c)$ which is not tangent to
any curve ${\cal F}(k,l)$ at this point.
An explicit formula for ${\sum_{j=1}^{\infty}}ch(M_{h,c}^{(j)})$
shows that  theorem A is true up to level $~n+1$ for the module $M_{h,c}$.
Here it is.
By $I$ let us denote  the set ${\bf N}$ for case
${\bf III_{-}^{\mbox {\small {\bf 00}}}}$
and the set $\{ 1,...,s \}$ for case
${\bf III_{+}^{\mbox {\small {\bf 00}}}}$. Then we distinguish two cases:

{\quad} {\quad} a) $c{\neq}1,{\,}25,{\,}24h+1$:
%% FOLLOWING LINE CANNOT BE BROKEN BEFORE 80 CHAR
$${\sum_{j=1}^{\infty}}ch(M_{h,c}^{(j)})=2({\sum_{j{\in}I}}ch(M_{h-k_il_i,c})).$$

{\quad} {\quad} b) $c=1~~\mbox{or}~~25~~\mbox{or}~~24h+1$:
$${\sum_{j=1}^{\infty}}ch(M_{h,c}^{(j)})={\sum_{j{\in}I}}ch(M_{h-k_il_i,c}).$$

These formulas follow immediately from the
Kac determinant formula and property 3 of
Jantzen's filtration.

$\beta )$ $~n+1=k_jl_j$ for some $j{\in}{\bf N}$ (or $1{\leq}j{\leq}s$).
Then we take ${\cal C}={\cal F}(k_j,l_j)$. This curve ${\cal C}$ is given by
the parametric equations $h=h(t),~~c=c(t)$ (see the
formulas in section 3). Corollary 3.2
shows that the module $M_{h(t),c(t)}$ has a
singular vector at level $n+1=k_jl_j$
 and this singular vector generates Verma submodule $M_{h(t)-n-1,c(t)}$.
Let $L(t){\stackrel {def}=}M_{h(t),c(t)}/M_{h(t)-n-1,c(t)}$.
The contravariant form {\bf B}:$M_{h(t),c(t)}{\rightarrow}{\bar M}_{h(t),c(t)}$
 vanishes on the
submodule $M_{h(t)-n-1,c(t)}$ and since it is a symmetric bilinear form,
it defines a form ${\bf {\tilde B}}:~L(t){\rightarrow}{\bar L}(t)$
with the same properties.
So we can speak about
Jantzen's filtration on $L(t)$ along this curve ${\cal C}$.
Unfortunately, we do not know the
determinant formula for module $L(t)$. Nevertheless, we can
calculate the following sum:
$${\sum_{t{\in}{\bf C}}}{\sum_{i=1}^{\infty}}q^{-h(t)}ch(L(t)^{(i)}).$$
This is going to be our first calculation.

The second calculation is described below.
It is easy to see from the
definition of Jantzen's filtration that if some submodule
$N{\subset}M_{h(t_0),c(t_0)}^{(k)}$ then the
image $\bar N$ of $N$ under projection
$M_{h(t_0),c(t_0)}{\rightarrow}L(t_0)$ is contained in $L(t_0)^{(k)}$ {\it
i.e.},
$\bar N{\subset}L(t_0)^{(k)}$. Since
theorems A, B and C are true up to level $n$
 we have some information about
Jantzen's filtration. Exactly, we know that
for $i<j$  Verma module generated by the singular vector at
level $k_il_i$ is contained in $M_{h(t_0),c(t_0)}^{(2i)}$
 ({\it i.e.} $M_{h(t_0)-k_il_i,c(t_0)}{\subset}M_{h(t_0),c(t_0)}^{(2i)}$).
Therefore, its image
${\bar M}_{h(t_0)-k_il_i,c(t_0)}{\subset}L(t_0)^{(2i)}$.
This gives us a
low boundary for
${\sum_{i=1}^{\infty}}q^{-h(t_0)}ch(L(t_0)^{(i)})$.
 Summing over all $t{\in}{\bf C}$ (using
the fact that in all other cases
except ${\bf III_{+}^{\mbox {\small {\bf 00}}}}$ and
 ${\bf III_{-}^{\mbox {\small {\bf 00}}}}$
we know  Jantzen's filtration exactly)
we obtain the  estimate from below
 for ${\sum_{t{\in}{\bf C}}}{\sum_{i=1}^{\infty}}q^{-h(t)}ch(L(t)^{(i)}).$
The remarkable fact is that these
two calculations give us the same answer. Thus
 we know Jantzen's filtration on $L(t)$
along this curve ${\cal C}$ for all
$t{\in}{\bf C}$. We immediately see that
the maximal submodule in $M_{h,c}$ is generated
by the singular vector at level $k_1l_1$ up to
level $n+1$. This finishes the proof.

\subsection{\bf First calculation.}

 This calculation is exactly the
first calculation from the paper by Feigin and Fuchs ({\bf [Fe-Fu 1]}).
We present it here in greater details for the sake of completeness.

  Let us take the compactification of
curve ${\cal F}(k_j,l_j)$ {\it i.e.} ${\bf CP}^1$. Then we have a
trivial vector bundle over it
$M_{h(t),c(t)}={{\bf U}({\mbox {\gotb n}}^-)}$ which is a direct
sum of trivial bundles
${{\bf U}({\mbox {\gotb n}}^-)}_i$ for $i{\in}{\bf Z}_+$. We have
subbundle ${\bf N}(t)$ which is generated by
singular vector ${\bf S}_{k_j,l_j}(t)$ over
Virasoro. Certainly, we have decompositions into the direct sums,
$$M_{h(t),c(t)}={{\bf U}({\mbox {\gotb n}}^-)}={\bigoplus_{i=0}^{\infty}}
{{\bf U}({\mbox {\gotb n}}^-)}_i~~~~~~\mbox{and}$$
$${\bf N}(t)={\bigoplus_{i=k_jl_j}^{\infty}}{\bf N}(t)_i.$$
Let us denote by $\eta$
the line bundle ${\bf N}(t)_{n+1}$ (remember that $n+1=k_jl_j $).
It is easy to see that ${\bf N}(t)_i=p(i-k_jl_j){\eta}$
for all $i{\in}{\bf Z}_+$,
where $p(i)$ is a partition function (we set $p(i)=0$ for $i<0$).
$$L(t)=M_{h(t),c(t)}/{\bf  N}(t),~~~~\mbox{in~~particular}$$
$$L(t)_i={{\bf U}({\mbox {\gotb n}}^-)}_i/{\bf N}(t)_i.$$
$\det ({\bf {\tilde  B}}_i)$ is a section of the following line bundle
$({\Lambda}^{\dim{\,}L(t)_i}L(t)_i)^{\otimes 2})'$ (it is obvious that
 ${\dim~L(t)_i}=p(i)-p(i-k_jl_j)$).
 This section is regular outside $0$ and ${\infty}$.
Let us denote by $P_i(0)$ and $P_i({\infty})$ the orders of
the poles of the section at zero
and infinity respectively. We have the following  formula
$${\sum_{t{\in}{\bf C}}}{\sum_{m=1}^{\infty}}\dim(L(t)^{(m)}_i)=
Eu(({\Lambda}^{\dim~L(t)_i}L(t)_i)^{\otimes 2})')+P_i(0)+P_i({\infty}),$$
where $Eu(\bullet)$ denotes the Euler number of the
vector bundle. This formula follows  from
property 3 of Jantzen's filtration. Therefore,
 we must calculate  the Euler number
and the  orders of the poles at zero and infinity of
the corresponding vector bundle. Let us calculate the
Euler number first.

{\bf Lemma 7.1. (see {\bf [Fe-Fu 1]})}
\begin {it}

The Euler number, $Eu(\eta)=k_j+l_j-2k_jl_j$.
\end {it}

{\bf Proof.} From  theorem
%3.8.
3.4. we see that line bundle $\eta$ has  section
${\bf S}_{k_j,l_j}(t)$ that has no zero and has two poles of orders
 $k_j(l_j-1)$ and $l_j(k_j-1)$.

 \quad \quad \quad \quad \quad \quad \quad \quad \quad \quad \quad \quad \quad
\quad \quad \quad \quad \quad \quad \quad \quad \quad        {\bf  Q.E.D.}

To calculate $Eu(({\Lambda}^{\dim~L(t)_i}L(t)_i)^{\otimes 2})')$ is
the same as to calculate the
 first Chern class, $\bf c_1$, of this vector bundle.
$${\bf c_1}(({\Lambda}^{\dim~L(t)_i}L(t)_i)^{\otimes 2})')=$$
$$=-2{\bf c_1}(({\Lambda}^{\dim~L(t)_i}L(t)_i))=-2{\bf c_1}(L(t)_i)=2{\bf
c_1}({\bf N}(t)_i)=2p(i-k_jl_j){\bf c_1}(\eta).$$

{\bf Lemma 7.2. (see [Fe-Fu 1])}
\begin {it}

The Euler number, $Eu(({\Lambda}^{\dim~L(t)_i}L(t)_i)^{\otimes 2})')
=2p(i-k_jl_j)(k_j+l_j-2k_jl_j)$
{}.
\end {it}

{\bf Proof.}  Obvious.

 \quad \quad \quad \quad \quad \quad \quad \quad \quad \quad \quad \quad \quad
\quad \quad \quad \quad \quad \quad \quad \quad \quad        {\bf  Q.E.D.}

Now let us calculate the numbers $P_i(0)$ and $P_i({\infty})$.
For example, near infinity we
can calculate the determinant of
the form ${\bf {\tilde  B}}_i$ as the determinant of
the principal minor of the matrix of the
contravariant form corresponding to the
following  part of
the basis $L_{-i_r}L_{-i_{r-1}}...L_{-i_2}L_{-i_1}(L_{k_j})^s$ where
$s<l_j$, $i_r{\geq}i_{r-1}{\geq}...{\geq}i_2{\geq}i_1{\geq}1$
and $i_m{\neq}k_j$
for all $m$. For $t{\rightarrow}{\infty}$
we know that $h{\sim}{{1-k_j^2}\over 4}t$
and $c{\sim}6t$. One can see that the
degree  in $t$ of the determinant of this minor
can be calculated as a sum of the
degrees of its diagonal entries (other products
have smaller degree).
The computation of this sum is similar to the computations in section 3
. We must take into account only that  $L_iL_{-i}{\bf v}_{h(t),c(t)}=
[-2ih(t)-{(i^3-i){\over}12}c(t)]{\bf v}_{h(t),c(t)}{\sim}
-{{i(i^2-k_j^2)}{\over}2}t{\bf v}_{h(t),c(t)}$
 which has degree 1 in $t$ if $i{\neq}k_j$ and that
$(L_{k_j})^s(L_{-k_j})^s{\bf v}_{h(t),c(t)}{\sim}
{(-{k \over {12}})^s}{\prod_{m=1}^s}
(k_j(13k_j-12(l_j+2s-2m))-1){\neq}0$ has degree 0 in $t$.
So $P_i({\infty})$ equals  the number
of all elements not equal to $k_j$ of all
partitions of $i$ which contain $k_j$ less then
$l_j$ times. A similar statement is true for $P_i(0)$.
We must only replace $k_j$ by $l_j$
and vice versa.
We obtain the following formulas:
$${\sum_{m=0}^{\infty}}P_m(\infty)u^m=
p(u)(1-u^{k_jl_j})\left(s(u)-{{u^{k_j}}\over {1-u^{k_j}}}\right),$$
$${\sum_{m=0}^{\infty}}P_m(0)u^m=
p(u)(1-u^{k_jl_j})\left(s(u)-{{u^{l_j}}\over {1-u^{l_j}}}\right),$$
where  $p(u)={\sum_{m=0}^{\infty}}p(m)u^m$,
$s(u)={\sum_{m=0}^{\infty}}{\sum_{s=0}^{\infty}}
u^{ms}$. Indeed,  $p(u){{u^m}\over {1-u^m}}$ is
a generating function for the number of all
elements equal to $~m$ of all partitions of positive integer.
Thus we proved the following proposition.

{\bf Proposition 7.3. (see [Fe-Fu 1])}
\begin {it}

$${\sum_{t{\in}{\bf C}}}
{\sum_{m=1}^{\infty}}q^{h(t)}ch(L(t)^{(m)})(q^{-1})=p(q)(1-q^{k_jl_j})
\left(2s(q)-{{u^{k_j}}\over {1-u^{k_j}}}-\right.$$
$$\left.-{{u^{l_j}}\over {1-u^{l_j}}}\right)-(2k_jl_j-k_j-l_j)p(q)q^{k_jl_j}.$$
\end {it}

{\bf Remark.} We have $n+1=k_jl_j$.

\subsection{\bf Second calculation.}

  This calculation is completely combinatorial and quite lengthy. It is a good
exercise in combinatorics to do it by yourself. In any case,
 everyone who is interested
in it can find it in [Fe-Fu] Chapter 2, paragraph 1, section 4.
The result is the same as the right hand side in the proposition 7.3. .

\subsection{\bf Final remarks.}

Comparison of these two calculations finishes the proof of
theorems A, B and~C.

One can see that  in
cases ${\bf III_{+}^{\mbox {\small {\bf 00}}}}$ and
 ${\bf III_{-}^{\mbox {\small {\bf 00}}}}$ the proof uses
the asymptotics of the formulas for the  singular
 vectors. It seems interesting to me to find another
 proof which does not use such kind of
information.

\section {On the structure of Verma modules over Neveu-Schwarz algebra.}

\subsection{\bf Notation.}

    Neveu-Schwarz is a Lie superalgebra with the
basis $L_i,L_{i+{1\over 2}}$ and $C$
where $i{\in}{\bf Z}$ and the following commutators:
$$  [L_i,C]=0,~~~~~[L_{i+{1\over 2}},C]=0 $$
$$ [L_i,L_j]=(j-i)L_{i+j} +{\delta_{-i,j}}{(j^3-j)\over 12}C,$$
$$ [L_{m+{1\over 2}},L_{n+{1\over 2}}]=
2L_{n+m+1}+{\delta_{0,m+n+1}}{(4n^2-1)\over 12}C,$$
$$ [L_{m+{1\over 2}},L_n]=({(n-1)\over 2}-m)L_{m+n+{1\over 2}}.$$

It is  $\bf {1\over 2}Z$-graded:
$\deg L_i=i,~\deg L_{i+{1\over 2}}=i+{1\over 2}$  and
 $\deg C=0.$
Let us denote by {\gotb h} the
Lie algebra with the basis $L_0~ and ~C$,  by {\gotb n$^-$}
the Lie algebra with the
basis $\{L_{-{i\over 2}},~~ i{\in}{\bf N}\}$ and by {\gotb n$^+$}
 the Lie algebra
with the basis $\{L_{-{i\over 2}},~~ i{\in}{\bf N}\}$.
We also denote by {\gotb b$^+$} the Lie
algebra with
the  basis  \{$L_{{i\over 2}}$and $C,~ i{\in}{\bf Z_+}$\}.
All these algebras  {\gotb h},
 {\gotb n$^-$}, {\gotb n$^+$} and  {\gotb b$^+$} are subalgebras of {\NeV}.
We have a Cartan type decomposition of {\NeV}

$$\quad\quad\quad {\NV}=
{{\mbox{\gotb n}}^-}{\oplus}{\mbox{\gotb h}}{\oplus}{{\mbox{\gotb n}}^+}$$
$$\quad\quad\quad {\NV}= {{\mbox{\gotb n}}^-}{\oplus}{{\mbox{\gotb b}}^+}.$$

Let $h,c{\in}\bf C$. Let's consider one dimensional module $\bf C{_{\em h,c}}$
over    {\gotb b$^+$} such that  {\gotb n$^+$} acts by zero, $L_0$ is
a multiplication by $h$ and $C$ is a multiplication by $c$.
Verma module $M_{h,c}$ over Neveu-Schwarz is (by definition) the induced
 module from $\bf C{_{\em h,c}}$

  $${\quad\quad\quad M_{h,c}={\mit Ind}^{\Vi}_{{\mbox{\gotb b}}^+}}
\bf C_{h,c}.$$

We have a natural inclusion of
${\bf C}_{h,c}{\hookrightarrow}{M_{h,c}}$. Therefore, we have
a vector ${\bf v}{\in}M_{h,c}$ corresponding
to ${\bf 1}{\in}{\bf C_{\em h,c}}$.
Sometimes we denote it by ${\bf v_{\em h,c}}$ to stress
that this vector lies in ${M_{h,c}}$.
Vector {\bf v} is called
the vacuum vector.

Let us make some remarks about Verma modules. First,
 any Verma module ${M_{h,c}}$
 is a free module
over {\bf U}({\gotb n$^-$}). We have the following basis in ${M_{h,c}}$:
  $${L_{-{i_k\over 2}}}{L_{-{i_{k-1}\over 2}}}...
{L_{-{i_2\over 2}}}{L_{-{i_1\over 2}}}{\bf v_{\em h,c}},$$
where ${i_k}{\geq}{i_{k-1}}{\geq}...{\geq}{i_1}{\geq}1$ and $i_j$ is
either odd or the multiple of 4 for any $j$.

The operator ${L_0}$ on $M_{h,c}$ is semisimple.
We can consider the eigenspace decomposition of $M_{h,c}$,
$${M_{h,c}=}{\bigoplus_{i=0}^{\infty}}~{M_{h,c}^{i\over 2}},$$
where ${L_0}$ acts as a
multiplication by ${h-{i\over 2}}$ on ${M_{h,c}^{i\over 2}}$.
It is easy to see that this decomposition respects
the grading on \NeV. We say that a vector ${\bf w}{\in}{M_{h,c}}$
has level {\em n} if ${\bf w}{\in}{M_{h,c}^n}$.

We call vector {\bf w} singular if it has some level {\em n}
(${\em n{\in}{\bf {1\over 2}Z_+}}$)
 and {\gotb n$^+$} acts by zero on this vector.
It is obvious that any singular vector generates
a submodule isomorphic to Verma module.
If a singular vector has level {\em n} then it generates $M_{h-n,c}$.

In the same way as for Virasoro
one can define a contavariant Hermitian form ${\bf B}(h,c)$
on Verma module $M_{h,c}$. One can see that the
spaces ${M_{h,c}^{i\over 2}}$, $i{\in}{{\bf Z}_{+}}$,
are orthogonal for different $i$.
We denote the restriction of the form ${\bf B}(h,c)$
to ${M_{h,c}^{i\over 2}}$ by ${\bf B}_{i\over 2}(h,c)$.

We have the following determinant formula (see [Kac-Wa]):

$$  {\det}^2({\bf B_n})={\mit Const}{\prod_{
\begin{array}{c}
{k,l{\in}{{\bf Z}_{+}}}\\
{\em kl}{\leq}{\bf n}\\
{k=l ~\mod(2)}
\end{array}}}
{\bf {\tilde  {\Phi}}}_{\em k,l}(h,c)^{{\tilde p}({n-kl)}\over 2}
{}~,~~~~~~~~~~~\mbox{where}$$

$$ {\bf {\tilde  {\Phi}}}_{\em k,l}(h,c)=
(h+{(k^2-1)c\over 24}+{5(1-k^2)-4(1-kl)\over 16}){\times}$$
$${\times}(h+{(l^2-1)c\over 24}+{5(1-l^2)-4(1-kl)\over 16})
+{{(k^2-l^2)}^2\over 64}.$$

The curve $ {\bf {\tilde  {\Phi}}}_{k,l}(h,c)=0$
can be given by the following formulas:

$$h={{1-k^2}\over 8}t+{{1-kl}\over 4}+{{1-l^2}\over 8}t^{-1}$$
$$c=3t+{15\over 2}+3t^{-1},~~~~~~~~~~~~~~\mbox{where} ~~t{\in}{\bf C}^*.$$

In order to formulate the main result, let us make the following substitution:
$$c=3{{(2p+q)(2q+p)}\over {2pq}},$$
$$h={{(p+q)^2-m^2}\over {8pq}}.$$

Then as in the case of Virasoro we have a quadruple of lines in
the plane (k,l):
$$m=pk+ql,~~m=pl+qk,~~0=m+pk+ql,~~0=m+pl+qk.$$

Let us choose one of these lines, for example $0=m+pk+ql$,
and  denote it by $l_{h,c}$.

\subsection{\bf On the structure of Verma Modules.}

We shall distinguish the following cases.

\begin{itemize}
\item[{\bf Case I.}] The line $l_{h,c}$ contains no integral points.

\item[{\bf Case II.}] The line $l_{h,c}$ contains exactly one integral point
$(k,l)$. We have the following subcases:

\begin{itemize}
\item[${\bf II_{-}.}$] The product $kl<0$.

\item[${\bf II_{0}.}$] The product $kl=0$.

\item[${\bf II_{+}.}$] We distinguish two cases:

\begin{itemize}
\item[{\bf a)}] $k=l ~\mod(2)$.

\item[{\bf b)}] $k{\neq}l ~\mod(2)$.
\end{itemize}
\end{itemize}

\item[{\bf Case III.}] The line $l_{h,c}$ contains infinitely
many integral points. Let us consider two neighboring
integral points on the line. Then the second
point can be obtained by adding some vector ${\bf v}$ to the first point.
We can write the vector ${\bf v}$ in coordinates $(v_1,v_2)$.
Numbers $v_1~$ and $~v_2$ are integers and we are
going to distinguish the following
two cases (we denote them by A and B respectively)

\begin{itemize}
\item[A)] $v_1~$ and $~v_2$ are both odd

\item[B)] one of $v_1~$ and $~v_2$ is odd and the other one is even

$v_1~$ and $~v_2$ can not be even at the same time since we
took neighboring points.
\end{itemize}

\item[${\bf Subcase ~~c{\leq}{3\over 2}}$.]

\item[${\bf III_{-}^{00}.}$] Line $l_{h,c}$ intersects
both axes at integral points.

\item[${\bf III_{-}^{0}.}$]  Line $l_{h,c}$ intersects  only
one axis at integral  point. Let $(k_1,l_1)$, $(k_2,l_2)$, $(k_3,l_3),
{\,}.{\,}.{\,}.$
be all integral points $(k,l)$ on the line $l_{h,c}$
 up to equivalence $~(k,l){\sim}(k^{'},l^{'})$~~iff~$kl=k^{'}l^{'}$
and such that $kl>0$. We ordered them in such a way that
${k_i}{l_i}<{k_{i+1}}{l_{i+1}}$ for
all $i{\in}{\bf N}$.

\begin{itemize}
\item[A)]  $v_1~~and~~v_2$ are both odd.

Then we distinguish two cases:

\begin{itemize}
\item[$\alpha$)]  $k_1{\neq}l_1~~\mod(2)$

\item[$\beta$)]   $k_1=l_1~~\mod(2)$
\end{itemize}

\item[B)]  one of $v_1~$ and $~v_2$ is odd
and the other one is even

We have four cases:

\begin{itemize}
\item[$\alpha$)] $k_1=l_1~~\mbox{and}~~k_2=l_2~~\mod(2)$

\item[$\beta$)] $k_1{\neq}l_1~~\mbox{and}~~k_2{\neq}l_2~~\mod(2)$
then $k_3=l_3~~\mbox{and}~~k_4=l_4~~\mod(2)$

\item[$\gamma$)] $k_1{\neq}l_1~~\mod(2)$,
$k_2=l_2~~\mbox{and}~~k_3=l_3~~\mod(2)$
then
\hfill $k_4{\neq}l_4~~\mod(2)$

\item[$\delta$)] $k_1=l_1~~\mbox{and}~~k_4=l_4~~\mod(2)$
then $k_2{\neq}l_2~~\mbox{and}~~k_3{\neq}l_3~~\mod(2)$
\end{itemize}
\end{itemize}

\item[${\bf III_{-}.}$]   Line $l_{h,c}$ intersects
 both axes at non-integral points.   Let $(k_1,l_1)$,
$(k_2,l_2)$, $(k_3,l_3),{\,}.{\,}.{\,}.$
be all integral points $(k,l)$ on the line $l_{h,c}$ such that
$k_i=l_i~~\mod (2)$ for all $i{\in}{\bf N}$
 up to equivalence $(k,l){\sim}(k^{'},l^{'})~$ iff $kl=k^{'}l^{'}$
 and such that $kl>0$. We ordered them in such a way
that ${k_i}{l_i}<{k_{i+1}}{l_{i+1}}$ for
all $i{\in}{\bf N}$.
In this case we draw an auxiliary line, $l_{h,c}^{'}$, parallel
to $l_{h,c}$ through the point $(k_1,-l_1)$.
  Let $(k^{'}_1,l^{'}_1)$, $(k^{'}_2,l^{'}_2)$,
$(k^{'}_3,l^{'}_3),{\,}.{\,}.{\,}.$
be all integral points $(k,l)$ on the line $l_{h,c}^{'}$ such that
$k^{'}_i=l^{'}_i~~\mod (2)$ for all $i{\in}{\bf N}$
 up to equivalence $(k,l){\sim}(k^{``},l^{``})~$ iff $kl=k^{``}l^{``}$
and such that $kl>0$.  We ordered them in such a
way that ${k^{'}_i}{l^{'}_i}<{k^{'}_{i+1}}{l^{'}_{i+1}}$ for
all $i{\in}{\bf N}$.
 Then it is easy to see that we have  the following inequalities:
$${k_1}{l_1}<{k_2}{l_2}<{k_1}{l_1}+{k_1^{'}}{l_1^{'}}<
{k_1}{l_1}+{k_2^{'}}{l_2^{'}}<
{k_3}{l_3}<{k_4}{l_4}<$$
$$<{k_1}{l_1}+{k_3^{'}}{l_3^{'}}<{k_1}{l_1}+{k_4^{'}}{l_4^{'}}<
{k_5}{l_5}<{k_6}{l_6}<...~~ .$$

We distinguish two case:

\begin{itemize}
\item[A)]  $~~~v_1~$ and $~v_2~$ are both odd.
Then either both sets
$\{{\,}(k_1,l_1)$, $(k_2,l_2)$, $(k_3,l_3),$ ${\,}.{\,}.{\,}.{\,}\}$
and $\{{\,}(k^{'}_1,l^{'}_1)$, $(k^{'}_2,l^{'}_2)$,
$(k^{'}_3,l^{'}_3),{\,}.{\,}.{\,}.{\,}\}$
are empty (subcase $\alpha$) or not
(subcase $\beta$).

\item[B)]  one of $v_1~$ and $~v_2$ is odd
and the other one is  even.
\end{itemize}

\item[${\bf Subcase ~~c{\geq}{{27}\over 2}}$.]

\item[${\bf III_{+}^{00}.}$] Line $l_{h,c}$ intersects
 both axes at integral points.

\item[${\bf III_{+}^{0}.}$]  Line $l_{h,c}$ intersects
  only one axis at  integral point.
Let $\{{\,}(k_1,l_1)$, $(k_2,l_2),$ ${\,}.{\,}.{\,}.{\,},(k_s,l_s)\}$ be all
 integral points $(k,l)$ on the line $l_{h,c}$ up to
equivalence $(k,l){\sim}(k^{'},l^{'})~$ iff $kl=k^{'}l^{'}$ and
such that $kl>0$.  We ordered them in such a way that
${k_i}{l_i}<{k_{i+1}}{l_{i+1}}$ for all $i{\in}{\{}1,2,...,s-1{\}}$.

\begin{itemize}
\item[A)]  $v_1~$ and $~v_2$ are both odd.

We distinguish two cases:

\begin{itemize}
\item[$\alpha$)]  $k_1{\neq}l_1~~ \mod(2)$

\item[$\beta$)]   $k_1=l_1~~\mod(2)$
\end{itemize}

\item[B)]  one of $v_1~$ and $~v_2$ is odd and
the other one is even

Then we have four cases:

\begin{itemize}
\item[$\alpha$)] $k_1=l_1~~\mbox{and}~~k_2=l_2~~\mod(2)$

\item[$\beta$)] $k_1{\neq}l_1~~\mbox{and}~~k_2{\neq}l_2~~\mod(2)$
then $k_3=l_3~~\mbox{and}~~k_4=l_4~~\mod(2)$

\item[$\gamma$)] $k_1{\neq}l_1~~\mod(2)$,
$k_2=l_2~~\mbox{and}~~k_3=l_3~~\mod(2)$ then $k_4{\neq}l_4~~\mod(2)$

\item[$\delta$)] $k_1=l_1~~\mbox{and}~~k_4=l_4~~\mod(2)$
then $k_2{\neq}l_2~~\mbox{and}~~k_3{\neq}l_3~~\mod(2)$
\end{itemize}
\end{itemize}

\item[${\bf III_{+}.}$]   Line $l_{h,c}$ intersects
 both axes at non-integral points.
Let $\{{\,}(k_1,l_1)$, $(k_2,l_2),$ ${\,}.{\,}.{\,}.{\,},(k_s,l_s)\}$
be all integral
points $(k,l)$ on the line $l_{h,c}$ such that $k_i=l_i~~\mod (2)$
 for all $i{\in}{\{}1,2,...,s{\}}$
up to equivalence $(k,l){\sim}(k^{'},l^{'})~$ iff $kl=k^{'}l^{'}$ and
such that $kl>0$.  We ordered them in such a way that
${k_i}{l_i}<{k_{i+1}}{l_{i+1}}$ for all $i{\in}{\{}1,2,...,s-1{\}}$.
In this case we draw an auxiliary line, $l_{h,c}^{'}$, parallel
to $l_{h,c}$ through the point $(k_1,-l_1)$.
 Let $(k^{'}_1,l^{'}_1)$, $(k^{'}_2,l^{'}_2),$ ${\,}.{\,}.{\,}.{\,},
(k^{'}_s,l^{'}_s)$
be all integral points $(k,l)$ on the line $l_{h,c}^{'}$ such that
$k^{'}_i=l^{'}_i~~\mod (2)$ for all $i{\in{\{}1,2,...,s-1{\}}}$
 up to equivalence $(k,l){\sim}(k^{``},l^{``})~$ iff $kl=k^{``}l^{``}$
and such that $kl>0$.  We ordered them in such a way that
${k^{'}_i}{l^{'}_i}<{k^{'}_{i+1}}{l^{'}_{i+1}}$
for $i{\in}{\{}1,2,...,s-1{\}}$.

 Then it is easy to see that we have  the following inequalities:
$${k_1}{l_1}<{k_2}{l_2}<{k_1}{l_1}+{k_1^{'}}{l_1^{'}}<
{k_1}{l_1}+{k_2^{'}}{l_2^{'}}<
{k_3}{l_3}<{k_4}{l_4}<$$
$$<{k_1}{l_1}+{k_3^{'}}{l_3^{'}}<{k_1}{l_1}+{k_4^{'}}{l_4^{'}}<
{k_5}{l_5}<{k_6}{l_6}<{\,}.{\,}.{\,}.~~ .$$

We distinguish two cases:
\begin{itemize}

\item[A)]  $v_1$ and $v_2$ are both  odd.
Then either  both sets
$\{(k_1,l_1)$, $(k_2,l_2)$, $(k_3,l_3),{\,}.{\,}.{\,}.\}$ and
$\{(k^{'}_1,l^{'}_1)$, $(k^{'}_2,l^{'}_2)$,
$(k^{'}_3,l^{'}_3),{\,}.{\,}.{\,}.\}$
are empty (subcase $\alpha$) or
 not (subcase $\beta$).

\item[B)]  one of $v_1~$ and $~v_2$ is odd
and the other one is even.
\end{itemize}
\end{itemize}

{{\bf Theorem D.}
\begin {it}

{\bf a)} In cases
${\bf I,II,III_{-},III^{0}_{-},III_{+}}$ and $\bf III^{0}_{+}$
all submodules of Verma module are generated by singular vectors.

{\bf b)} \quad i) In cases
${\bf I,~II_{-},II_{0},II_{+}.a),~III^{0}_{-}.A)
\alpha),~III^{0}_{+}.A)\alpha),~III_{-}.A)\alpha) }$
and ${\bf III_{+}.A)\alpha)}$
 Verma module is irreducible.

\quad \quad ii) In case $\bf II_{+}.b)$ Verma module
$M_{h,c}$ has a unique submodule generated
by the singular vector at  level $kl$. This submodule is isomorphic
to Verma module $M_{h-kl,c}$ which
is irreducible (case $\bf II_{-}$).

\quad \quad iii) In  cases $\bf III^{0}_{-}.A)\beta)
,~III^{0}_{-}.B)$ and $\bf III_{-}$  we have an infinite number
 of singular vectors. All singular vectors and relations
between them are shown on the diagrams below.
Singular vectors are denoted by points with
 their weights indicated. An arrow or a chain
of arrows from one point to another
means that the second singular vector
lies in the submodule generated by the first
one.

\quad \quad i{\rm v})  In cases $\bf III^{0}_{+}.A)\beta)
,~III^{0}_{+}.B)$ and $\bf III_{+}$  we have a finite number
 of singular vectors (maybe zero). All singular vectors and relations
between them are shown on the diagrams below.
Singular vectors are denoted by points with
 their weights indicated. An arrow or a chain
of arrows from one point to another
means that the second singular  vector lies
in the submodule generated by the first one.
\end {it}

\subsection{\bf Sketch of the proof.}

Let ${M_{h,c}}$ be a Verma module with the central charge $c$ and
the highest weight $h$.
Then the following theorem can be proved
similarly to theorem {\bf 3.1.}.

{\bf Theorem 8.1.}
\begin{it}
  At each level $n{\in}{1\over 2}{\bf Z}$ only one
singular vector {\bf w} can exist.
If the singular vector, {\bf w}, exists
then it is given by the following formula

$${\bf w}={({L_{-{1\over 2}}})^{2n}}{\bf v_{\em h,c}}+$$
$$+{\sum_{\begin{array}{c}{i_k}+...+{i_1}=2n\\{i_k}{\geq}{i_{k-1}}{\geq}...
{\geq}{i_1}{\geq}1\\{i_k{\geq}3}\\where~i_j~ is ~either ~odd\\ ~or~a~
multiple ~of ~4 ~for ~any ~j\end{array}}}{{\bf P}^{(n)}_{{i_1},...,{i_k}}}
(h,c){L_{-{{i_k}\over 2}}}{L_{-{{i_{k-1}}\over 2}}}...{L_{-{{i_2}\over 2}}}
{L_{-{{i_1}\over 2}}}{\bf v_{\em h,c}}$$
which defines {\bf w} up to multiplication by a constant.
${{\bf P}^{(n)}_{{i_1},...,{i_k}}}(h,c)$ are polynomials in  h  and  c.
\end{it}

One can see that the proof of the structure of Verma modules in the simple
cases  uses  only the Kac determinant formula and theorem {\bf 3.1}.
We have the  analog of both {\it i.e.}, theorem {\bf 8.1.} and
the determinant formula.
Repeating the arguments which we used in the
case of Virasoro (almost word for
word with minor modifications ) one can see that
{\bf Theorem  D} is true.

\pagebreak
\setlength{\unitlength}{0.012500in}%
\begin{picture}(355,285)( 0,530)
\thicklines
\put( 60,665){\circle*{10}}
\put( 60,710){\circle*{10}}
\put( 60,755){\circle*{10}}
\put(235,785){\circle*{10}}
\put(200,750){\circle*{10}}
\put(200,710){\circle*{10}}
\put(200,670){\circle*{10}}
\put(200,630){\circle*{10}}
\put(200,590){\circle*{10}}
\put(270,750){\circle*{10}}
\put(270,710){\circle*{10}}
\put(270,670){\circle*{10}}
\put(270,630){\circle*{10}}
\put(270,590){\circle*{10}}
\put( 60,750){\vector( 0,-1){ 35}}
\put( 60,705){\vector( 0,-1){ 35}}
\put( 60,660){\vector( 0,-1){ 35}}
\put(200,705){\vector( 0,-1){ 30}}
\put(270,705){\vector( 0,-1){ 30}}
\put(265,705){\vector(-2,-1){ 60}}
\put(205,705){\vector( 2,-1){ 60}}
\put(200,665){\vector( 0,-1){ 30}}
\put(270,665){\vector( 0,-1){ 30}}
\put(200,745){\line( 0,-1){ 25}}
\put(200,720){\vector( 0,-1){  5}}
\put(270,745){\vector( 0,-1){ 30}}
\put(265,745){\vector(-2,-1){ 60}}
\put(205,745){\vector( 2,-1){ 60}}
\put(240,780){\vector( 1,-1){ 25}}
\put(230,780){\vector(-1,-1){ 25}}
\put(205,665){\vector( 2,-1){ 60}}
\put(265,665){\vector(-2,-1){ 60}}
\put(200,625){\vector( 0,-1){ 30}}
\put(270,625){\line( 0,-1){ 25}}
\put(270,600){\line( 0, 1){  0}}
\put(270,600){\vector( 0,-1){  5}}
\put(205,625){\vector( 2,-1){ 60}}
\put(265,625){\vector(-2,-1){ 60}}
\put(200,585){\line( 0,-1){ 10}}
\put(270,585){\line( 0,-1){ 10}}
\put( 50,610){\makebox(0,0)[lb]{\smash{\Large {\bf .}}}}
\put( 60,610){\makebox(0,0)[lb]{\smash{\Large {\bf .}}}}
\put( 70,610){\makebox(0,0)[lb]{\smash{\Large {\bf .}}}}
\put( 15,810){\makebox(0,0)[lb]{\smash{\Large {$\bf Case~~~III^{\mbox
 {\small {\bf 0}}}_{\bf -}{\bf .}{\rm A)\beta)}
{}~~~~~~~Case~~~~III_{\bf -}$}}}}
\put(210,570){\makebox(0,0)[lb]{\smash{\Large {\bf   .   .   .}}}}
\put(185,785){\makebox(0,0)[lb]{\smash{\small {\bf (h , c)}}}}
\put( 15,755){\makebox(0,0)[lb]{\smash{\small {\bf (h , c)}}}}
\put(85,625){\makebox(0,0)[lb]{\smash{\small {$\bf
(h-k_{1}l_{1}-k^{'}_{3}l^{'}_{3},c)$}}}}
\put(85,705){\makebox(0,0)[lb]{\smash{\small {$\bf
(h-k_{1}l_{1}-k^{'}_{1}l^{'}_{1},c)$}}}}
\put(280,710){\makebox(0,0)[lb]{\smash{\small {$\bf
(h-k_{1}l_{1}-k^{'}_{2}l^{'}_{2},c)$}}}}
\put(280,630){\makebox(0,0)[lb]{\smash{\small {$\bf
(h-k_{1}l_{1}-k^{'}_{4}l^{'}_{4},c)$}}}}
\put(110,665){\makebox(0,0)[lb]{\smash{\small {$\bf (h-k_{3}l_{3},c)$}}}}
\put(110,750){\makebox(0,0)[lb]{\smash{\small {$\bf (h-k_{1}l_{1},c)$}}}}
\put(110,590){\makebox(0,0)[lb]{\smash{\small {$\bf (h-k_{5}l_{5},c)$}}}}
\put(285,750){\makebox(0,0)[lb]{\smash{\small {$\bf (h-k_{2}l_{2},c)$}}}}
\put(285,670){\makebox(0,0)[lb]{\smash{\small {$\bf (h-k_{4}l_{4},c)$}}}}
\put(285,590){\makebox(0,0)[lb]{\smash{\small {$\bf (h-k_{6}l_{6},c)$}}}}
\put(-15,710){\makebox(0,0)[lb]{\smash{\small {$\bf (h-k_{1}l_{1},c)$}}}}
\put(-15,660){\makebox(0,0)[lb]{\smash{\small {$\bf (h-k_{2}l_{2},c)$}}}}
\end{picture}

\setlength{\unitlength}{0.012500in}%
\begin{picture}(290,247)(0,600)
\thicklines
\put( 70,775){\circle*{10}}
\put( 70,745){\circle*{10}}
\put( 70,715){\circle*{10}}
\put( 70,605){\circle*{10}}
\put( 70,635){\circle*{10}}
\put(230,790){\circle*{10}}
\put(205,760){\circle*{10}}
\put(255,760){\circle*{10}}
\put(205,730){\circle*{10}}
\put(255,730){\circle*{10}}
\put(205,700){\circle*{10}}
\put(255,700){\circle*{10}}
\put(205,670){\circle*{10}}
\put(255,670){\circle*{10}}
\put(220,650){\circle*{2}}
\put(230,650){\circle*{2}}
\put(240,650){\circle*{2}}
\put( 60,680){\circle*{2}}
\put( 60,680){\circle*{2}}
\put( 60,680){\circle*{2}}
\put( 70,680){\circle*{2}}
\put( 80,680){\circle*{2}}
\put(205,615){\circle*{10}}
\put(255,615){\circle*{10}}
\put(230,585){\circle*{10}}
\put( 70,770){\vector( 0,-1){ 20}}
\put( 70,740){\vector( 0,-1){ 20}}
\put( 70,630){\vector( 0,-1){ 20}}
\put( 70,710){\line( 0,-1){ 10}}
\put( 70,660){\vector( 0,-1){ 20}}
\put(225,785){\vector(-1,-1){ 20}}
\put(235,785){\vector( 1,-1){ 20}}
\put(205,755){\vector( 0,-1){ 20}}
\put(255,755){\vector( 0,-1){ 20}}
\put(210,755){\vector( 2,-1){ 40}}
\put(250,755){\vector(-2,-1){ 40}}
\put(205,725){\vector( 0,-1){ 20}}
\put(255,725){\vector( 0,-1){ 20}}
\put(250,725){\vector(-2,-1){ 40}}
\put(250,755){\vector(-2,-1){ 40}}
\put(250,755){\vector(-2,-1){ 40}}
\put(210,725){\vector( 2,-1){ 40}}
\put(250,695){\vector(-2,-1){ 40}}
\put(210,695){\vector( 2,-1){ 40}}
\put(205,695){\vector( 0,-1){ 20}}
\put(255,695){\vector( 0,-1){ 20}}
\put(205,665){\line( 0,-1){ 10}}
\put(255,665){\line( 0,-1){ 10}}
\put(205,690){\line( 0, 1){  0}}
\put(205,690){\line( 0, 1){  0}}
\put(205,640){\vector( 0,-1){ 20}}
\put(230,640){\vector(-1,-1){ 20}}
\put(255,640){\vector( 0,-1){ 20}}
\put(230,640){\vector( 1,-1){ 20}}
\put(250,610){\vector(-1,-1){ 20}}
\put(210,610){\vector( 1,-1){ 20}}
\put( 20,825){\makebox(0,0)[lb]{\smash{\Large {$\bf Case~~~III^{\mbox {\small
{\bf 0}}}_{\bf +}{\bf .}{\rm A)\beta)}$}}}}
\put(200,825){\makebox(0,0)[lb]{\smash{\Large {$\bf Case ~~~III_{\bf +}$}}}}
\put( 15,770){\makebox(0,0)[lb]{\smash{\small {$\bf (h,c)$}}}}
\put( -5,740){\makebox(0,0)[lb]{\smash{\small {$\bf (h-k_{1}l_{1},c)$}}}}
\put( -5,715){\makebox(0,0)[lb]{\smash{\small {$\bf (h-k_{2}l_{2},c)$}}}}
\put(-30,630){\makebox(0,0)[lb]{\smash{\small {$\bf (h-k_{s-1}l_{s-1},c)$}}}}
\put( -5,600){\makebox(0,0)[lb]{\smash{\small {$\bf (h-k_{s}l_{s},c)$}}}}
\put(130,755){\makebox(0,0)[lb]{\smash{\small {$\bf (h-k_{1}l_{1},c)$}}}}
\put( 90,725){\makebox(0,0)[lb]{\smash{\small {$\bf
(h-k_{1}l_{1}-k^{'}_{1}l^{'}_{1},c)$}}}}
\put(130,695){\makebox(0,0)[lb]{\smash{\small {$\bf (h-k_{3}l_{3},c)$}}}}
\put( 90,660){\makebox(0,0)[lb]{\smash{\small {$\bf
(h-k_{1}l_{1}-k^{'}_{3}l^{'}_{3},c)$}}}}
\put(165,785){\makebox(0,0)[lb]{\smash{\small {$\bf (h,c)$}}}}
\put(265,755){\makebox(0,0)[lb]{\smash{\small {$\bf (h-k_{2}l_{2},c)$}}}}
\put(265,725){\makebox(0,0)[lb]{\smash{\small {$\bf
(h-k_{1}l_{1}-k^{'}_{2}l^{'}_{2},c)$}}}}
\put(265,695){\makebox(0,0)[lb]{\smash{\small {$\bf (h-k_{4}l_{4},c)$}}}}
\put(265,665){\makebox(0,0)[lb]{\smash{\small {$\bf
(h-k_{1}l_{1}-k^{'}_{4}l^{'}_{4},c)$}}}}
\end{picture}

\pagebreak

\setlength{\unitlength}{0.012500in}%
\begin{picture}(355,285)( 0,480)
\thicklines
\put( 60,665){\circle*{10}}
\put( 60,710){\circle*{10}}
\put( 60,755){\circle*{10}}
\put( 60,620){\circle*{10}}
\put( 60,575){\circle*{10}}
\put(235,785){\circle*{10}}
\put(200,750){\circle*{10}}
\put(200,710){\circle*{10}}
\put(200,670){\circle*{10}}
\put(200,630){\circle*{10}}
\put(200,590){\circle*{10}}
\put(270,750){\circle*{10}}
\put(270,710){\circle*{10}}
\put(270,670){\circle*{10}}
\put(270,630){\circle*{10}}
\put(270,590){\circle*{10}}
\put( 60,750){\vector( 0,-1){ 35}}
\put( 60,705){\vector( 0,-1){ 35}}
\put( 60,660){\vector( 0,-1){ 35}}
\put( 60,615){\vector( 0,-1){ 35}}
\put( 60,570){\vector( 0,-1){ 35}}
\put(200,705){\vector( 0,-1){ 30}}
\put(270,705){\vector( 0,-1){ 30}}
\put(265,705){\vector(-2,-1){ 60}}
\put(205,705){\vector( 2,-1){ 60}}
\put(200,665){\vector( 0,-1){ 30}}
\put(270,665){\vector( 0,-1){ 30}}
\put(200,745){\line( 0,-1){ 25}}
\put(200,720){\vector( 0,-1){  5}}
\put(270,745){\vector( 0,-1){ 30}}
\put(265,745){\vector(-2,-1){ 60}}
\put(205,745){\vector( 2,-1){ 60}}
\put(240,780){\vector( 1,-1){ 25}}
\put(230,780){\vector(-1,-1){ 25}}
\put(205,665){\vector( 2,-1){ 60}}
\put(265,665){\vector(-2,-1){ 60}}
\put(200,625){\vector( 0,-1){ 30}}
\put(270,625){\line( 0,-1){ 25}}
\put(270,600){\line( 0, 1){  0}}
\put(270,600){\vector( 0,-1){  5}}
\put(205,625){\vector( 2,-1){ 60}}
\put(265,625){\vector(-2,-1){ 60}}
\put(200,585){\line( 0,-1){ 10}}
\put(270,585){\line( 0,-1){ 10}}
\put( 50,520){\makebox(0,0)[lb]{\smash{\Large {\bf .}}}}
\put( 60,520){\makebox(0,0)[lb]{\smash{\Large {\bf .}}}}
\put( 70,520){\makebox(0,0)[lb]{\smash{\Large {\bf .}}}}
\put( 15,810){\makebox(0,0)[lb]{\smash{\Large {$\bf Case~~~III^{\mbox
 {\small {\bf 0}}}_{\bf -}{\bf .}{\rm B)\beta)}
{}~~~~Case~~~III^{\mbox
 {\small {\bf 0}}}_{\bf -}{\bf .}{\rm B)\alpha)}$}}}}
\put(210,570){\makebox(0,0)[lb]{\smash{\Large {\bf   .   .   .}}}}
\put(185,785){\makebox(0,0)[lb]{\smash{\small {\bf (h , c)}}}}
\put( 15,755){\makebox(0,0)[lb]{\smash{\small {\bf (h , c)}}}}
\put(110,625){\makebox(0,0)[lb]{\smash{\small {$\bf (h-k_{7}l_{7},c)$}}}}
\put(110,705){\makebox(0,0)[lb]{\smash{\small {$\bf (h-k_{3}l_{3},c)$}}}}
\put(285,710){\makebox(0,0)[lb]{\smash{\small {$\bf (h-k_{4}l_{4},c)$}}}}
\put(285,630){\makebox(0,0)[lb]{\smash{\small {$\bf (h-k_{8}l_{8},c)$}}}}
\put(110,665){\makebox(0,0)[lb]{\smash{\small {$\bf (h-k_{5}l_{5},c)$}}}}
\put(110,750){\makebox(0,0)[lb]{\smash{\small {$\bf (h-k_{1}l_{1},c)$}}}}
\put(110,590){\makebox(0,0)[lb]{\smash{\small {$\bf (h-k_{9}l_{9},c)$}}}}
\put(285,750){\makebox(0,0)[lb]{\smash{\small {$\bf (h-k_{2}l_{2},c)$}}}}
\put(285,670){\makebox(0,0)[lb]{\smash{\small {$\bf (h-k_{6}l_{6},c)$}}}}
\put(285,590){\makebox(0,0)[lb]{\smash{\small {$\bf (h-k_{10}l_{10},c)$}}}}
\put(-15,710){\makebox(0,0)[lb]{\smash{\small {$\bf (h-k_{3}l_{3},c)$}}}}
\put(-15,660){\makebox(0,0)[lb]{\smash{\small {$\bf (h-k_{4}l_{4},c)$}}}}
\put(-15,610){\makebox(0,0)[lb]{\smash{\small {$\bf (h-k_{7}l_{7},c)$}}}}
\put(-15,560){\makebox(0,0)[lb]{\smash{\small {$\bf (h-k_{8}l_{8},c)$}}}}
\end{picture}

\setlength{\unitlength}{0.012500in}%
\begin{picture}(355,285)( 0,550)
\thicklines
\put( 60,665){\circle*{10}}
\put( 60,710){\circle*{10}}
\put( 60,755){\circle*{10}}
\put( 60,620){\circle*{10}}
\put( 60,575){\circle*{10}}
\put(235,785){\circle*{10}}
\put(200,750){\circle*{10}}
\put(200,710){\circle*{10}}
\put(200,670){\circle*{10}}
\put(200,630){\circle*{10}}
\put(200,590){\circle*{10}}
\put(270,750){\circle*{10}}
\put(270,710){\circle*{10}}
\put(270,670){\circle*{10}}
\put(270,630){\circle*{10}}
\put(270,590){\circle*{10}}
\put( 60,750){\vector( 0,-1){ 35}}
\put( 60,705){\vector( 0,-1){ 35}}
\put( 60,660){\vector( 0,-1){ 35}}
\put( 60,615){\vector( 0,-1){ 35}}
\put( 60,570){\vector( 0,-1){ 35}}
\put(200,705){\vector( 0,-1){ 30}}
\put(270,705){\vector( 0,-1){ 30}}
\put(265,705){\vector(-2,-1){ 60}}
\put(205,705){\vector( 2,-1){ 60}}
\put(200,665){\vector( 0,-1){ 30}}
\put(270,665){\vector( 0,-1){ 30}}
\put(200,745){\line( 0,-1){ 25}}
\put(200,720){\vector( 0,-1){  5}}
\put(270,745){\vector( 0,-1){ 30}}
\put(265,745){\vector(-2,-1){ 60}}
\put(205,745){\vector( 2,-1){ 60}}
\put(240,780){\vector( 1,-1){ 25}}
\put(230,780){\vector(-1,-1){ 25}}
\put(205,665){\vector( 2,-1){ 60}}
\put(265,665){\vector(-2,-1){ 60}}
\put(200,625){\vector( 0,-1){ 30}}
\put(270,625){\line( 0,-1){ 25}}
\put(270,600){\line( 0, 1){  0}}
\put(270,600){\vector( 0,-1){  5}}
\put(205,625){\vector( 2,-1){ 60}}
\put(265,625){\vector(-2,-1){ 60}}
\put(200,585){\line( 0,-1){ 10}}
\put(270,585){\line( 0,-1){ 10}}
\put( 50,520){\makebox(0,0)[lb]{\smash{\Large {\bf .}}}}
\put( 60,520){\makebox(0,0)[lb]{\smash{\Large {\bf .}}}}
\put( 70,520){\makebox(0,0)[lb]{\smash{\Large {\bf .}}}}
\put( 15,810){\makebox(0,0)[lb]{\smash{\Large {$\bf Case~~~III^{\mbox
 {\small {\bf 0}}}_{\bf -}{\bf .}{\rm B)\delta)}
{}~~~~Case~~~III^{\mbox
 {\small {\bf 0}}}_{\bf -}{\bf .}{\rm B)\gamma)}$}}}}
\put(210,570){\makebox(0,0)[lb]{\smash{\Large {\bf   .   .   .}}}}
\put(185,785){\makebox(0,0)[lb]{\smash{\small {\bf (h , c)}}}}
\put( 15,755){\makebox(0,0)[lb]{\smash{\small {\bf (h , c)}}}}
\put(110,625){\makebox(0,0)[lb]{\smash{\small {$\bf (h-k_{8}l_{8},c)$}}}}
\put(110,705){\makebox(0,0)[lb]{\smash{\small {$\bf (h-k_{4}l_{4},c)$}}}}
\put(285,710){\makebox(0,0)[lb]{\smash{\small {$\bf (h-k_{5}l_{5},c)$}}}}
\put(285,630){\makebox(0,0)[lb]{\smash{\small {$\bf (h-k_{9}l_{9},c)$}}}}
\put(110,665){\makebox(0,0)[lb]{\smash{\small {$\bf (h-k_{6}l_{6},c)$}}}}
\put(110,750){\makebox(0,0)[lb]{\smash{\small {$\bf (h-k_{2}l_{2},c)$}}}}
\put(105,590){\makebox(0,0)[lb]{\smash{\small {$\bf (h-k_{10}l_{10},c)$}}}}
\put(285,750){\makebox(0,0)[lb]{\smash{\small {$\bf (h-k_{3}l_{3},c)$}}}}
\put(285,670){\makebox(0,0)[lb]{\smash{\small {$\bf (h-k_{7}l_{7},c)$}}}}
\put(285,590){\makebox(0,0)[lb]{\smash{\small {$\bf (h-k_{11}l_{11},c)$}}}}
\put(-15,710){\makebox(0,0)[lb]{\smash{\small {$\bf (h-k_{1}l_{1},c)$}}}}
\put(-15,660){\makebox(0,0)[lb]{\smash{\small {$\bf (h-k_{4}l_{4},c)$}}}}
\put(-15,610){\makebox(0,0)[lb]{\smash{\small {$\bf (h-k_{5}l_{5},c)$}}}}
\put(-15,560){\makebox(0,0)[lb]{\smash{\small {$\bf (h-k_{8}l_{8},c)$}}}}
\end{picture}

\pagebreak

\setlength{\unitlength}{0.012500in}%
\begin{picture}(290,247)(0,560)
\thicklines
\put( 70,775){\circle*{10}}
\put( 70,745){\circle*{10}}
\put( 70,715){\circle*{10}}
\put( 70,685){\circle*{10}}
\put( 70,655){\circle*{10}}
\put( 70,575){\circle*{10}}
\put(230,790){\circle*{10}}
\put(205,760){\circle*{10}}
\put(255,760){\circle*{10}}
\put(205,730){\circle*{10}}
\put(255,730){\circle*{10}}
\put(205,700){\circle*{10}}
\put(255,700){\circle*{10}}
\put(205,670){\circle*{10}}
\put(255,670){\circle*{10}}
\put(220,650){\circle*{2}}
\put(230,650){\circle*{2}}
\put(240,650){\circle*{2}}
\put( 60,620){\circle*{2}}
\put( 70,620){\circle*{2}}
\put( 80,620){\circle*{2}}
\put(205,615){\circle*{10}}
\put(255,615){\circle*{10}}
\put(230,585){\circle*{10}}
\put( 70,770){\vector( 0,-1){ 20}}
\put( 70,740){\vector( 0,-1){ 20}}
\put( 70,680){\vector( 0,-1){ 20}}
\put( 70,710){\vector( 0,-1){ 20}}
\put( 70,650){\line( 0,-1){ 10}}
\put( 70,600){\vector( 0,-1){ 20}}
\put(225,785){\vector(-1,-1){ 20}}
\put(235,785){\vector( 1,-1){ 20}}
\put(205,755){\vector( 0,-1){ 20}}
\put(255,755){\vector( 0,-1){ 20}}
\put(210,755){\vector( 2,-1){ 40}}
\put(250,755){\vector(-2,-1){ 40}}
\put(205,725){\vector( 0,-1){ 20}}
\put(255,725){\vector( 0,-1){ 20}}
\put(250,725){\vector(-2,-1){ 40}}
\put(250,755){\vector(-2,-1){ 40}}
\put(250,755){\vector(-2,-1){ 40}}
\put(210,725){\vector( 2,-1){ 40}}
\put(250,695){\vector(-2,-1){ 40}}
\put(210,695){\vector( 2,-1){ 40}}
\put(205,695){\vector( 0,-1){ 20}}
\put(255,695){\vector( 0,-1){ 20}}
\put(205,665){\line( 0,-1){ 10}}
\put(255,665){\line( 0,-1){ 10}}
\put(205,690){\line( 0, 1){  0}}
\put(205,690){\line( 0, 1){  0}}
\put(205,640){\vector( 0,-1){ 20}}
\put(230,640){\vector(-1,-1){ 20}}
\put(255,640){\vector( 0,-1){ 20}}
\put(230,640){\vector( 1,-1){ 20}}
\put(250,610){\vector(-1,-1){ 20}}
\put(210,610){\vector( 1,-1){ 20}}
\put( 20,825){\makebox(0,0)[lb]{\smash{\Large {$\bf Case~~~III^{\mbox {\small
{\bf 0}}}_{\bf +}{\bf .}{\rm B)\beta)}$}}}}
\put(200,825){\makebox(0,0)[lb]{\smash{\Large {$\bf Case~~~III^{\mbox {\small
{\bf 0}}}_{\bf +}{\bf .}{\rm B)\alpha)}$}}}}
\put( 15,770){\makebox(0,0)[lb]{\smash{\small {$\bf (h,c)$}}}}
\put( -5,740){\makebox(0,0)[lb]{\smash{\small {$\bf (h-k_{3}l_{3},c)$}}}}
\put( -5,710){\makebox(0,0)[lb]{\smash{\small {$\bf (h-k_{4}l_{4},c)$}}}}
\put( -5,680){\makebox(0,0)[lb]{\smash{\small {$\bf (h-k_{7}l_{7},c)$}}}}
\put( -5,650){\makebox(0,0)[lb]{\smash{\small {$\bf (h-k_{8}l_{8},c)$}}}}
\put(130,755){\makebox(0,0)[lb]{\smash{\small {$\bf (h-k_{1}l_{1},c)$}}}}
\put(130,725){\makebox(0,0)[lb]{\smash{\small {$\bf (h-k_{3}l_{3},c)$}}}}
\put(130,695){\makebox(0,0)[lb]{\smash{\small {$\bf (h-k_{5}l_{5},c)$}}}}
\put(130,660){\makebox(0,0)[lb]{\smash{\small {$\bf (h-k_{7}l_{7},c)$}}}}
\put(165,785){\makebox(0,0)[lb]{\smash{\small {$\bf (h,c)$}}}}
\put(265,755){\makebox(0,0)[lb]{\smash{\small {$\bf (h-k_{2}l_{2},c)$}}}}
\put(265,725){\makebox(0,0)[lb]{\smash{\small {$\bf (h-k_{4}l_{4},c)$}}}}
\put(265,695){\makebox(0,0)[lb]{\smash{\small {$\bf (h-k_{6}l_{6},c)$}}}}
\put(265,665){\makebox(0,0)[lb]{\smash{\small {$\bf (h-k_{8}l_{8},c)$}}}}
\end{picture}

\setlength{\unitlength}{0.012500in}%
\begin{picture}(290,247)(0,630)
\thicklines
\put( 70,775){\circle*{10}}
\put( 70,745){\circle*{10}}
\put( 70,715){\circle*{10}}
\put( 70,685){\circle*{10}}
\put( 70,655){\circle*{10}}
\put( 70,575){\circle*{10}}
\put(230,790){\circle*{10}}
\put(205,760){\circle*{10}}
\put(255,760){\circle*{10}}
\put(205,730){\circle*{10}}
\put(255,730){\circle*{10}}
\put(205,700){\circle*{10}}
\put(255,700){\circle*{10}}
\put(205,670){\circle*{10}}
\put(255,670){\circle*{10}}
\put(220,650){\circle*{2}}
\put(230,650){\circle*{2}}
\put(240,650){\circle*{2}}
\put( 60,620){\circle*{2}}
\put( 70,620){\circle*{2}}
\put( 80,620){\circle*{2}}
\put(205,615){\circle*{10}}
\put(255,615){\circle*{10}}
\put(230,585){\circle*{10}}
\put( 70,770){\vector( 0,-1){ 20}}
\put( 70,740){\vector( 0,-1){ 20}}
\put( 70,680){\vector( 0,-1){ 20}}
\put( 70,710){\vector( 0,-1){ 20}}
\put( 70,650){\line( 0,-1){ 10}}
\put( 70,600){\vector( 0,-1){ 20}}
\put(225,785){\vector(-1,-1){ 20}}
\put(235,785){\vector( 1,-1){ 20}}
\put(205,755){\vector( 0,-1){ 20}}
\put(255,755){\vector( 0,-1){ 20}}
\put(210,755){\vector( 2,-1){ 40}}
\put(250,755){\vector(-2,-1){ 40}}
\put(205,725){\vector( 0,-1){ 20}}
\put(255,725){\vector( 0,-1){ 20}}
\put(250,725){\vector(-2,-1){ 40}}
\put(250,755){\vector(-2,-1){ 40}}
\put(250,755){\vector(-2,-1){ 40}}
\put(210,725){\vector( 2,-1){ 40}}
\put(250,695){\vector(-2,-1){ 40}}
\put(210,695){\vector( 2,-1){ 40}}
\put(205,695){\vector( 0,-1){ 20}}
\put(255,695){\vector( 0,-1){ 20}}
\put(205,665){\line( 0,-1){ 10}}
\put(255,665){\line( 0,-1){ 10}}
\put(205,690){\line( 0, 1){  0}}
\put(205,690){\line( 0, 1){  0}}
\put(205,640){\vector( 0,-1){ 20}}
\put(230,640){\vector(-1,-1){ 20}}
\put(255,640){\vector( 0,-1){ 20}}
\put(230,640){\vector( 1,-1){ 20}}
\put(250,610){\vector(-1,-1){ 20}}
\put(210,610){\vector( 1,-1){ 20}}
\put( 20,825){\makebox(0,0)[lb]{\smash{\Large {$\bf Case~~~III^{\mbox {\small
{\bf 0}}}_{\bf +}{\bf .}{\rm B)\delta)}$}}}}
\put(200,825){\makebox(0,0)[lb]{\smash{\Large {$\bf Case~~~III^{\mbox {\small
{\bf 0}}}_{\bf +}{\bf .}{\rm B)\gamma)}$}}}}
\put( 15,770){\makebox(0,0)[lb]{\smash{\small {$\bf (h,c)$}}}}
\put( -5,740){\makebox(0,0)[lb]{\smash{\small {$\bf (h-k_{1}l_{1},c)$}}}}
\put( -5,710){\makebox(0,0)[lb]{\smash{\small {$\bf (h-k_{4}l_{4},c)$}}}}
\put( -5,680){\makebox(0,0)[lb]{\smash{\small {$\bf (h-k_{5}l_{5},c)$}}}}
\put( -5,650){\makebox(0,0)[lb]{\smash{\small {$\bf (h-k_{8}l_{8},c)$}}}}
\put(130,755){\makebox(0,0)[lb]{\smash{\small {$\bf (h-k_{2}l_{2},c)$}}}}
\put(130,725){\makebox(0,0)[lb]{\smash{\small {$\bf (h-k_{4}l_{4},c)$}}}}
\put(130,695){\makebox(0,0)[lb]{\smash{\small {$\bf (h-k_{6}l_{6},c)$}}}}
\put(130,660){\makebox(0,0)[lb]{\smash{\small {$\bf (h-k_{8}l_{8},c)$}}}}
\put(165,785){\makebox(0,0)[lb]{\smash{\small {$\bf (h,c)$}}}}
\put(265,755){\makebox(0,0)[lb]{\smash{\small {$\bf (h-k_{3}l_{3},c)$}}}}
\put(265,725){\makebox(0,0)[lb]{\smash{\small {$\bf (h-k_{5}l_{5},c)$}}}}
\put(265,695){\makebox(0,0)[lb]{\smash{\small {$\bf (h-k_{7}l_{7},c)$}}}}
\put(265,665){\makebox(0,0)[lb]{\smash{\small {$\bf (h-k_{9}l_{9},c)$}}}}
\end{picture}

\pagebreak
\setlength{\unitlength}{0.012500in}%
\begin{picture}(355,285)(5,505)
\thicklines
\put(207,727){$\bullet$}
\put(222,757){$\bullet$}
\put(237,787){$\bullet$}
\put(252,817){$\bullet$}
\put(162,637){$\bullet$}
\put(147,607){$\bullet$}
\put(132,577){$\bullet$}
\put(180,670){\circle{5}}
\put(195,700){\circle{5}}
\put(180,560){\vector( 0, 1){280}}
\put(180,560){\line( 0,-1){  5}}
\put(  5,700){\vector( 1, 0){355}}
\put(195,705){\line( 0,-1){  5}}
\put(210,705){\line( 0,-1){  5}}
\put(225,705){\line( 0,-1){  5}}
\put(240,705){\line( 0,-1){  5}}
\put(255,705){\line( 0,-1){  5}}
\put(270,705){\line( 0,-1){  5}}
\put(285,705){\line( 0,-1){  5}}
\put(300,705){\line( 0,-1){  5}}
\put(315,705){\line( 0,-1){  5}}
\put(330,705){\line( 0,-1){  5}}
\put(165,705){\line( 0,-1){  5}}
\put(150,705){\line( 0,-1){  5}}
\put(135,705){\line( 0,-1){  5}}
\put(120,705){\line( 0,-1){  5}}
\put(105,705){\line( 0,-1){  5}}
\put( 90,705){\line( 0,-1){  5}}
\put( 75,705){\line( 0,-1){  5}}
\put( 60,705){\line( 0,-1){  5}}
\put( 45,705){\line( 0,-1){  5}}
\put( 30,705){\line( 0,-1){  5}}
\put( 15,705){\line( 0,-1){  5}}
\put(180,715){\line( 1, 0){  5}}
\put(180,730){\line( 1, 0){  5}}
\put(180,745){\line( 1, 0){  5}}
\put(180,760){\line( 1, 0){  5}}
\put(180,775){\line( 1, 0){  5}}
\put(180,790){\line( 1, 0){  5}}
\put(180,805){\line( 1, 0){  5}}
\put(180,820){\line( 1, 0){  5}}
\put(180,685){\line( 1, 0){  5}}
\put(180,670){\line( 1, 0){  5}}
\put(180,655){\line( 1, 0){  5}}
\put(180,640){\line( 1, 0){  5}}
\put(180,625){\line( 1, 0){  5}}
\put(180,610){\line( 1, 0){  5}}
\put(180,595){\line( 1, 0){  5}}
\put(180,580){\line( 1, 0){  5}}
\put(180,565){\line( 1, 0){  5}}
\put( 10,790){\line( 0, 1){  0}}
\put( 10,790){\line( 0, 1){  0}}
\put( 10,795){\line( 1, 0){ 70}}
\put(240,795){\line( 0, 1){  0}}
\put(255,820){\line(-1,-2){ 15}}
\put(240,790){\line(-1,-2){ 15}}
\put(225,760){\line(-1,-2){ 15}}
\put(210,730){\line(-1,-2){ 15}}
\put(195,700){\line(-1,-2){ 15}}
\put(180,670){\line(-1,-2){ 15}}
\put(165,640){\line(-1,-2){ 15}}
\put(150,610){\line(-1,-2){ 15}}
\put(135,580){\line(-1,-2){ 10}}
\put(125,560){\line( 0, 1){  0}}
\put(255,820){\line( 1, 2){ 10}}
\put(210,690){\makebox(0,0)[lb]{\smash{\small 2}}}
\put(225,690){\makebox(0,0)[lb]{\smash{\small 3}}}
\put(240,690){\makebox(0,0)[lb]{\smash{\small 4}}}
\put(255,690){\makebox(0,0)[lb]{\smash{\small 5}}}
\put(270,690){\makebox(0,0)[lb]{\smash{\small 6}}}
\put(285,690){\makebox(0,0)[lb]{\smash{\small 7}}}
\put(300,690){\makebox(0,0)[lb]{\smash{\small 8}}}
\put(315,690){\makebox(0,0)[lb]{\smash{\small 9}}}
\put(330,690){\makebox(0,0)[lb]{\smash{\small 10}}}
\put(160,690){\makebox(0,0)[lb]{\smash{\small -1}}}
\put(145,690){\makebox(0,0)[lb]{\smash{\small -2}}}
\put(130,690){\makebox(0,0)[lb]{\smash{\small -3}}}
\put(115,690){\makebox(0,0)[lb]{\smash{\small -4}}}
\put(100,690){\makebox(0,0)[lb]{\smash{\small -5}}}
\put( 85,690){\makebox(0,0)[lb]{\smash{\small -6}}}
\put( 70,690){\makebox(0,0)[lb]{\smash{\small -7}}}
\put( 55,690){\makebox(0,0)[lb]{\smash{\small -8}}}
\put( 40,690){\makebox(0,0)[lb]{\smash{\small -9}}}
\put( 10,690){\makebox(0,0)[lb]{\smash{\small -11}}}
\put(170,685){\makebox(0,0)[lb]{\smash{\small -1}}}
\put(170,655){\makebox(0,0)[lb]{\smash{\small -3}}}
\put(170,640){\makebox(0,0)[lb]{\smash{\small -4}}}
\put(170,625){\makebox(0,0)[lb]{\smash{\small -5}}}
\put(170,610){\makebox(0,0)[lb]{\smash{\small -6}}}
\put(170,595){\makebox(0,0)[lb]{\smash{\small -7}}}
\put(170,580){\makebox(0,0)[lb]{\smash{\small -8}}}
\put(170,565){\makebox(0,0)[lb]{\smash{\small -9}}}
\put(170,820){\makebox(0,0)[lb]{\smash{\small 8}}}
\put(170,805){\makebox(0,0)[lb]{\smash{\small 7}}}
\put(170,790){\makebox(0,0)[lb]{\smash{\small 6}}}
\put(170,775){\makebox(0,0)[lb]{\smash{\small 5}}}
\put(170,760){\makebox(0,0)[lb]{\smash{\small 4}}}
\put(170,745){\makebox(0,0)[lb]{\smash{\small 3}}}
\put(170,730){\makebox(0,0)[lb]{\smash{\small 2}}}
\put(170,715){\makebox(0,0)[lb]{\smash{\small 1}}}
\put( 10,800){\makebox(0,0)[lb]{\smash{\Large {\bf Figure   1.}}}}
\put(220,725){\makebox(0,0)[lb]{\smash{\bf (k}}}
\put(235,720){\makebox(0,0)[lb]{\smash{\bf 1}}}
\put(245,725){\makebox(0,0)[lb]{\smash{\bf , l}}}
\put(260,720){\makebox(0,0)[lb]{\smash{\bf 1}}}
\put(270,725){\makebox(0,0)[lb]{\smash{\bf )}}}
\put(235,755){\makebox(0,0)[lb]{\smash{\bf (k}}}
\put(250,750){\makebox(0,0)[lb]{\smash{\bf 2}}}
\put(260,755){\makebox(0,0)[lb]{\smash{\bf , l}}}
\put(275,750){\makebox(0,0)[lb]{\smash{\bf 2}}}
\put(285,755){\makebox(0,0)[lb]{\smash{\bf )}}}
\put(250,785){\makebox(0,0)[lb]{\smash{\bf (k}}}
\put(265,780){\makebox(0,0)[lb]{\smash{\bf 3}}}
\put(275,785){\makebox(0,0)[lb]{\smash{\bf , l}}}
\put(290,780){\makebox(0,0)[lb]{\smash{\bf 3}}}
\put(300,785){\makebox(0,0)[lb]{\smash{\bf )}}}
\put(265,815){\makebox(0,0)[lb]{\smash{\bf (k}}}
\put(280,810){\makebox(0,0)[lb]{\smash{\bf 4}}}
\put(290,815){\makebox(0,0)[lb]{\smash{\bf , l}}}
\put(305,810){\makebox(0,0)[lb]{\smash{\bf 4}}}
\put(315,815){\makebox(0,0)[lb]{\smash{\bf )}}}
\put(165,670){\makebox(0,0)[lb]{\smash{\small -2}}}
\put(195,685){\makebox(0,0)[lb]{\smash{\small 1}}}
\end{picture}
\setlength{\unitlength}{0.012500in}%
\begin{picture}(355,274)(363,540)
\thicklines
\put(207,407){$\bullet$}
\put(237,452){$\bullet$}
\put(267,497){$\bullet$}
\put(147,317){$\bullet$}
\put(117,272){$\bullet$}
\put(180,365){\circle{5}}
\put(  5,394){\vector( 1, 0){355}}
\put(195,399){\line( 0,-1){  5}}
\put(210,399){\line( 0,-1){  5}}
\put(225,399){\line( 0,-1){  5}}
\put(240,399){\line( 0,-1){  5}}
\put(255,399){\line( 0,-1){  5}}
\put(270,399){\line( 0,-1){  5}}
\put(285,399){\line( 0,-1){  5}}
\put(300,399){\line( 0,-1){  5}}
\put(315,399){\line( 0,-1){  5}}
\put(330,399){\line( 0,-1){  5}}
\put(165,399){\line( 0,-1){  5}}
\put(150,399){\line( 0,-1){  5}}
\put(135,399){\line( 0,-1){  5}}
\put(120,399){\line( 0,-1){  5}}
\put(105,399){\line( 0,-1){  5}}
\put( 90,399){\line( 0,-1){  5}}
\put( 75,399){\line( 0,-1){  5}}
\put( 60,399){\line( 0,-1){  5}}
\put( 45,399){\line( 0,-1){  5}}
\put( 30,399){\line( 0,-1){  5}}
\put( 15,399){\line( 0,-1){  5}}
\put(180,409){\line( 1, 0){  5}}
\put(180,424){\line( 1, 0){  5}}
\put(180,439){\line( 1, 0){  5}}
\put(180,454){\line( 1, 0){  5}}
\put(180,469){\line( 1, 0){  5}}
\put(180,484){\line( 1, 0){  5}}
\put(180,499){\line( 1, 0){  5}}
\put(180,514){\line( 1, 0){  5}}
\put(180,379){\line( 1, 0){  5}}
\put(180,364){\line( 1, 0){  5}}
\put(180,349){\line( 1, 0){  5}}
\put(180,334){\line( 1, 0){  5}}
\put(180,319){\line( 1, 0){  5}}
\put(180,304){\line( 1, 0){  5}}
\put(180,289){\line( 1, 0){  5}}
\put(180,274){\line( 1, 0){  5}}
\put( 10,484){\line( 0, 1){  0}}
\put( 10,484){\line( 0, 1){  0}}
\put( 10,489){\line( 1, 0){ 70}}
\put(180,265){\vector( 0, 1){269}}
\put(270,500){\line(-2,-3){160}}
\put(270,500){\line( 2, 3){ 20}}
\put(195,374){\makebox(0,0)[lb]{\smash{\small 1}}}
\put(210,384){\makebox(0,0)[lb]{\smash{\small 2}}}
\put(225,384){\makebox(0,0)[lb]{\smash{\small 3}}}
\put(240,384){\makebox(0,0)[lb]{\smash{\small 4}}}
\put(255,384){\makebox(0,0)[lb]{\smash{\small 5}}}
\put(270,384){\makebox(0,0)[lb]{\smash{\small 6}}}
\put(285,384){\makebox(0,0)[lb]{\smash{\small 7}}}
\put(300,384){\makebox(0,0)[lb]{\smash{\small 8}}}
\put(315,384){\makebox(0,0)[lb]{\smash{\small 9}}}
\put(330,384){\makebox(0,0)[lb]{\smash{\small 10}}}
\put(160,384){\makebox(0,0)[lb]{\smash{\small -1}}}
\put(145,384){\makebox(0,0)[lb]{\smash{\small -2}}}
\put(130,384){\makebox(0,0)[lb]{\smash{\small -3}}}
\put(115,384){\makebox(0,0)[lb]{\smash{\small -4}}}
\put(100,384){\makebox(0,0)[lb]{\smash{\small -5}}}
\put( 85,384){\makebox(0,0)[lb]{\smash{\small -6}}}
\put( 70,384){\makebox(0,0)[lb]{\smash{\small -7}}}
\put( 55,384){\makebox(0,0)[lb]{\smash{\small -8}}}
\put( 40,384){\makebox(0,0)[lb]{\smash{\small -9}}}
\put( 10,384){\makebox(0,0)[lb]{\smash{\small -11}}}
\put(170,379){\makebox(0,0)[lb]{\smash{\small -1}}}
\put(170,349){\makebox(0,0)[lb]{\smash{\small -3}}}
\put(170,334){\makebox(0,0)[lb]{\smash{\small -4}}}
\put(170,319){\makebox(0,0)[lb]{\smash{\small -5}}}
\put(170,304){\makebox(0,0)[lb]{\smash{\small -6}}}
\put(170,289){\makebox(0,0)[lb]{\smash{\small -7}}}
\put(170,274){\makebox(0,0)[lb]{\smash{\small -8}}}
\put(170,514){\makebox(0,0)[lb]{\smash{\small 8}}}
\put(170,499){\makebox(0,0)[lb]{\smash{\small 7}}}
\put(170,484){\makebox(0,0)[lb]{\smash{\small 6}}}
\put(170,469){\makebox(0,0)[lb]{\smash{\small 5}}}
\put(170,454){\makebox(0,0)[lb]{\smash{\small 4}}}
\put(170,439){\makebox(0,0)[lb]{\smash{\small 3}}}
\put(170,424){\makebox(0,0)[lb]{\smash{\small 2}}}
\put(170,409){\makebox(0,0)[lb]{\smash{\small 1}}}
\put(165,364){\makebox(0,0)[lb]{\smash{\small -2}}}
\put(220,405){\makebox(0,0)[lb]{\smash{\bf (k}}}
\put(235,400){\makebox(0,0)[lb]{\smash{\bf 1}}}
\put(245,405){\makebox(0,0)[lb]{\smash{\bf , l}}}
\put(270,405){\makebox(0,0)[lb]{\smash{\bf )}}}
\put(135,315){\makebox(0,0)[lb]{\smash{\bf )}}}
\put(260,400){\makebox(0,0)[lb]{\smash{\bf 1}}}
\put(125,310){\makebox(0,0)[lb]{\smash{\bf 2}}}
\put(110,315){\makebox(0,0)[lb]{\smash{\bf , l}}}
\put(100,310){\makebox(0,0)[lb]{\smash{\bf 2}}}
\put( 85,315){\makebox(0,0)[lb]{\smash{\bf (k}}}
\put(250,450){\makebox(0,0)[lb]{\smash{\bf (k}}}
\put(265,445){\makebox(0,0)[lb]{\smash{\bf 3}}}
\put(275,450){\makebox(0,0)[lb]{\smash{\bf , l}}}
\put(290,445){\makebox(0,0)[lb]{\smash{\bf 3}}}
\put(300,450){\makebox(0,0)[lb]{\smash{\bf )}}}
\put( 55,270){\makebox(0,0)[lb]{\smash{\bf (k}}}
\put( 70,265){\makebox(0,0)[lb]{\smash{\bf 4}}}
\put( 80,270){\makebox(0,0)[lb]{\smash{\bf , l}}}
\put( 95,265){\makebox(0,0)[lb]{\smash{\bf 4}}}
\put(105,270){\makebox(0,0)[lb]{\smash{\bf )}}}
\put(280,495){\makebox(0,0)[lb]{\smash{\bf (k}}}
\put(295,490){\makebox(0,0)[lb]{\smash{\bf 5}}}
\put(305,495){\makebox(0,0)[lb]{\smash{\bf , l}}}
\put(320,490){\makebox(0,0)[lb]{\smash{\bf 5}}}
\put(330,495){\makebox(0,0)[lb]{\smash{\bf )}}}
\put( 90,495){\makebox(0,0)[lb]{\smash{\Large {\bf 2}}}}
\put( 10,495){\makebox(0,0)[lb]{\smash{\Large {\bf Figure}}}}
\put(100,495){\makebox(0,0)[lb]{\smash{\Large {\bf .}}}}
\end{picture}

\pagebreak
\setlength{\unitlength}{0.012500in}%
\begin{picture}(355,271)(5,510)
\thicklines
\put(192,707){$\bullet$}
\put(222,752){$\bullet$}
\put(252,797){$\bullet$}
\put(162,662){$\bullet$}
\put(132,617){$\bullet$}
\put(102,572){$\bullet$}
\put(252,767){$\bullet$}
\put(222,722){$\bullet$}
\put(195,680){\circle{5}}
\put(162,632){$\bullet$}
\put(132,587){$\bullet$}
\put(180,570){\vector( 0, 1){264}}
\put(  5,694){\vector( 1, 0){355}}
\put(195,699){\line( 0,-1){  5}}
\put(210,699){\line( 0,-1){  5}}
\put(225,699){\line( 0,-1){  5}}
\put(240,699){\line( 0,-1){  5}}
\put(255,699){\line( 0,-1){  5}}
\put(270,699){\line( 0,-1){  5}}
\put(285,699){\line( 0,-1){  5}}
\put(300,699){\line( 0,-1){  5}}
\put(315,699){\line( 0,-1){  5}}
\put(330,699){\line( 0,-1){  5}}
\put(165,699){\line( 0,-1){  5}}
\put(150,699){\line( 0,-1){  5}}
\put(135,699){\line( 0,-1){  5}}
\put(120,699){\line( 0,-1){  5}}
\put(105,699){\line( 0,-1){  5}}
\put( 90,699){\line( 0,-1){  5}}
\put( 75,699){\line( 0,-1){  5}}
\put( 60,699){\line( 0,-1){  5}}
\put( 45,699){\line( 0,-1){  5}}
\put( 30,699){\line( 0,-1){  5}}
\put( 15,699){\line( 0,-1){  5}}
\put(180,709){\line( 1, 0){  5}}
\put(180,724){\line( 1, 0){  5}}
\put(180,739){\line( 1, 0){  5}}
\put(180,754){\line( 1, 0){  5}}
\put(180,769){\line( 1, 0){  5}}
\put(180,784){\line( 1, 0){  5}}
\put(180,799){\line( 1, 0){  5}}
\put(180,814){\line( 1, 0){  5}}
\put(180,679){\line( 1, 0){  5}}
\put(180,664){\line( 1, 0){  5}}
\put(180,649){\line( 1, 0){  5}}
\put(180,634){\line( 1, 0){  5}}
\put(180,619){\line( 1, 0){  5}}
\put(180,604){\line( 1, 0){  5}}
\put(180,589){\line( 1, 0){  5}}
\put(180,574){\line( 1, 0){  5}}
\put( 10,784){\line( 0, 1){  0}}
\put( 10,784){\line( 0, 1){  0}}
\put( 10,789){\line( 1, 0){ 70}}
\put(277,833){\line(-2,-3){180}}
\put(285,820){\line( 0, 1){  0}}
\multiput(125,575)(15.77641,23.66461){10}{\line( 2, 3){  8.012}}
\put(210,684){\makebox(0,0)[lb]{\smash{\small 2}}}
\put(225,684){\makebox(0,0)[lb]{\smash{\small 3}}}
\put(240,684){\makebox(0,0)[lb]{\smash{\small 4}}}
\put(255,684){\makebox(0,0)[lb]{\smash{\small 5}}}
\put(270,684){\makebox(0,0)[lb]{\smash{\small 6}}}
\put(285,684){\makebox(0,0)[lb]{\smash{\small 7}}}
\put(300,684){\makebox(0,0)[lb]{\smash{\small 8}}}
\put(315,684){\makebox(0,0)[lb]{\smash{\small 9}}}
\put(330,684){\makebox(0,0)[lb]{\smash{\small 10}}}
\put(160,684){\makebox(0,0)[lb]{\smash{\small -1}}}
\put(145,684){\makebox(0,0)[lb]{\smash{\small -2}}}
\put(130,684){\makebox(0,0)[lb]{\smash{\small -3}}}
\put(115,684){\makebox(0,0)[lb]{\smash{\small -4}}}
\put(100,684){\makebox(0,0)[lb]{\smash{\small -5}}}
\put( 85,684){\makebox(0,0)[lb]{\smash{\small -6}}}
\put( 70,684){\makebox(0,0)[lb]{\smash{\small -7}}}
\put( 55,684){\makebox(0,0)[lb]{\smash{\small -8}}}
\put( 40,684){\makebox(0,0)[lb]{\smash{\small -9}}}
\put(170,664){\makebox(0,0)[lb]{\smash{\small -2}}}
\put(170,649){\makebox(0,0)[lb]{\smash{\small -3}}}
\put(170,634){\makebox(0,0)[lb]{\smash{\small -4}}}
\put(170,619){\makebox(0,0)[lb]{\smash{\small -5}}}
\put(170,604){\makebox(0,0)[lb]{\smash{\small -6}}}
\put(170,589){\makebox(0,0)[lb]{\smash{\small -7}}}
\put(170,574){\makebox(0,0)[lb]{\smash{\small -8}}}
\put(170,814){\makebox(0,0)[lb]{\smash{\small 8}}}
\put(170,799){\makebox(0,0)[lb]{\smash{\small 7}}}
\put(170,784){\makebox(0,0)[lb]{\smash{\small 6}}}
\put(170,769){\makebox(0,0)[lb]{\smash{\small 5}}}
\put(170,754){\makebox(0,0)[lb]{\smash{\small 4}}}
\put(170,739){\makebox(0,0)[lb]{\smash{\small 3}}}
\put(170,724){\makebox(0,0)[lb]{\smash{\small 2}}}
\put(170,709){\makebox(0,0)[lb]{\smash{\small 1}}}
\put( 10,684){\makebox(0,0)[lb]{\smash{\small -11}}}
\put(165,679){\makebox(0,0)[lb]{\smash{\small -1}}}
\put( 70,615){\makebox(0,0)[lb]{\smash{\bf (k}}}
\put( 95,615){\makebox(0,0)[lb]{\smash{\bf , l}}}
\put(120,615){\makebox(0,0)[lb]{\smash{\bf )}}}
\put( 40,570){\makebox(0,0)[lb]{\smash{\bf (k}}}
\put( 65,570){\makebox(0,0)[lb]{\smash{\bf , l}}}
\put( 90,570){\makebox(0,0)[lb]{\smash{\bf )}}}
\put(100,660){\makebox(0,0)[lb]{\smash{\bf (k}}}
\put(115,655){\makebox(0,0)[lb]{\smash{\bf 2}}}
\put(125,660){\makebox(0,0)[lb]{\smash{\bf , l}}}
\put(140,655){\makebox(0,0)[lb]{\smash{\bf 2}}}
\put(150,660){\makebox(0,0)[lb]{\smash{\bf )}}}
\put( 85,610){\makebox(0,0)[lb]{\smash{\bf 4}}}
\put(110,610){\makebox(0,0)[lb]{\smash{\bf 4}}}
\put( 55,565){\makebox(0,0)[lb]{\smash{\bf 6}}}
\put( 80,565){\makebox(0,0)[lb]{\smash{\bf 6}}}
\put( 10,800){\makebox(0,0)[lb]{\smash{\Large {\bf Figure   3.}}}}
\put(105,720){\makebox(0,0)[lb]{\smash{\bf (k}}}
\put(120,715){\makebox(0,0)[lb]{\smash{\bf 1}}}
\put(130,720){\makebox(0,0)[lb]{\smash{\bf , l}}}
\put(145,715){\makebox(0,0)[lb]{\smash{\bf 1}}}
\put(155,720){\makebox(0,0)[lb]{\smash{\bf )}}}
\put(185,785){\makebox(0,0)[lb]{\smash{\bf (k}}}
\put(200,780){\makebox(0,0)[lb]{\smash{\bf 3}}}
\put(210,785){\makebox(0,0)[lb]{\smash{\bf , l}}}
\put(225,780){\makebox(0,0)[lb]{\smash{\bf 3}}}
\put(235,785){\makebox(0,0)[lb]{\smash{\bf )}}}
\put(210,820){\makebox(0,0)[lb]{\smash{\bf (k}}}
\put(225,815){\makebox(0,0)[lb]{\smash{\bf 5}}}
\put(235,820){\makebox(0,0)[lb]{\smash{\bf , l}}}
\put(250,815){\makebox(0,0)[lb]{\smash{\bf 5}}}
\put(260,820){\makebox(0,0)[lb]{\smash{\bf )}}}
\put(265,765){\makebox(0,0)[lb]{\smash{\bf (k'}}}
\put(280,760){\makebox(0,0)[lb]{\smash{\bf 4}}}
\put(290,765){\makebox(0,0)[lb]{\smash{\bf , l'}}}
\put(305,760){\makebox(0,0)[lb]{\smash{\bf 4}}}
\put(315,765){\makebox(0,0)[lb]{\smash{\bf )}}}
\put(235,720){\makebox(0,0)[lb]{\smash{\bf (k'}}}
\put(260,720){\makebox(0,0)[lb]{\smash{\bf , l'}}}
\put(285,720){\makebox(0,0)[lb]{\smash{\bf )}}}
\put(195,630){\makebox(0,0)[lb]{\smash{\bf (k'}}}
\put(220,630){\makebox(0,0)[lb]{\smash{\bf , l'}}}
\put(245,630){\makebox(0,0)[lb]{\smash{\bf )}}}
\put(195,585){\makebox(0,0)[lb]{\smash{\bf (k'}}}
\put(220,585){\makebox(0,0)[lb]{\smash{\bf , l'}}}
\put(245,585){\makebox(0,0)[lb]{\smash{\bf )}}}
\put(210,625){\makebox(0,0)[lb]{\smash{\bf 1}}}
\put(235,625){\makebox(0,0)[lb]{\smash{\bf 1}}}
\put(250,715){\makebox(0,0)[lb]{\smash{\bf 2}}}
\put(275,715){\makebox(0,0)[lb]{\smash{\bf 2}}}
\put(210,580){\makebox(0,0)[lb]{\smash{\bf 3}}}
\put(235,580){\makebox(0,0)[lb]{\smash{\bf 3}}}
\end{picture}

\setlength{\unitlength}{0.012500in}%
\begin{picture}(350,268)(5,540)
\thicklines
\put( 40,720){\circle{5}}
\put( 77,697){$\bullet$}
\put(117,677){$\bullet$}
\put(157,657){$\bullet$}
\put(200,640){\circle{5}}
\put(240,620){\circle{5}}
\put(280,600){\circle{5}}
\put(320,580){\circle{5}}
\put( 40,565){\vector( 0, 1){265}}
\put( 40,760){\line( 1, 0){  5}}
\put( 20,580){\vector( 1, 0){335}}
\put( 40,820){\line( 1, 0){  5}}
\put( 40,700){\line( 1, 0){  5}}
\put( 40,640){\line( 1, 0){  5}}
\put(100,585){\line( 0,-1){  5}}
\put(100,580){\line( 0, 1){  5}}
\put( 40,600){\line( 1, 0){  5}}
\put( 40,620){\line( 1, 0){  5}}
\put( 40,660){\line( 1, 0){  5}}
\put( 40,680){\line( 1, 0){  5}}
\put( 40,720){\line( 1, 0){  5}}
\put( 40,740){\line( 1, 0){  5}}
\put( 40,780){\line( 1, 0){  5}}
\put( 40,800){\line( 1, 0){  5}}
\put(140,585){\line( 0,-1){  5}}
\put(140,580){\line( 0, 1){  5}}
\put( 60,585){\line( 0,-1){  5}}
\put( 60,580){\line( 0, 1){  5}}
\put( 80,585){\line( 0,-1){  5}}
\put( 80,580){\line( 0, 1){  5}}
\put(120,585){\line( 0,-1){  5}}
\put(120,580){\line( 0, 1){  5}}
\put(180,585){\line( 0,-1){  5}}
\put(180,580){\line( 0, 1){  5}}
\put(160,585){\line( 0,-1){  5}}
\put(160,580){\line( 0, 1){  5}}
\put(180,585){\line( 0,-1){  5}}
\put(180,580){\line( 0, 1){  5}}
\put(180,585){\line( 0,-1){  5}}
\put(180,580){\line( 0, 1){  5}}
\put(200,585){\line( 0,-1){  5}}
\put(200,580){\line( 0, 1){  5}}
\put(220,585){\line( 0,-1){  5}}
\put(220,580){\line( 0, 1){  5}}
\put(240,585){\line( 0,-1){  5}}
\put(240,580){\line( 0, 1){  5}}
\put(260,585){\line( 0,-1){  5}}
\put(260,580){\line( 0, 1){  5}}
\put(280,585){\line( 0,-1){  5}}
\put(280,580){\line( 0, 1){  5}}
\put(300,585){\line( 0,-1){  5}}
\put(300,580){\line( 0, 1){  5}}
\put(320,585){\line( 0,-1){  5}}
\put(320,580){\line( 0, 1){  5}}
\put(340,585){\line( 0,-1){  5}}
\put(340,580){\line( 0, 1){  5}}
\put( 20,580){\line(-1, 0){ 15}}
\put( 20,585){\line( 0,-1){  5}}
\put( 20,580){\line( 0, 1){  5}}
\put(220,800){\line( 1, 0){ 65}}
\put(285,800){\line( 1, 0){  5}}
\put(330,575){\line(-2, 1){310}}
\put( 30,600){\makebox(0,0)[lb]{\smash{\small 1}}}
\put( 30,620){\makebox(0,0)[lb]{\smash{\small 2}}}
\put( 30,640){\makebox(0,0)[lb]{\smash{\small 3}}}
\put( 30,660){\makebox(0,0)[lb]{\smash{\small 4}}}
\put( 30,680){\makebox(0,0)[lb]{\smash{\small 5}}}
\put( 30,700){\makebox(0,0)[lb]{\smash{\small 6}}}
\put( 30,720){\makebox(0,0)[lb]{\smash{\small 7}}}
\put( 30,740){\makebox(0,0)[lb]{\smash{\small 8}}}
\put( 30,760){\makebox(0,0)[lb]{\smash{\small 9}}}
\put( 24,780){\makebox(0,0)[lb]{\smash{\small 10}}}
\put( 24,800){\makebox(0,0)[lb]{\smash{\small 11}}}
\put( 24,820){\makebox(0,0)[lb]{\smash{\small 12}}}
\put( 60,570){\makebox(0,0)[lb]{\smash{\small 1}}}
\put( 15,570){\makebox(0,0)[lb]{\smash{\small -1}}}
\put( 80,570){\makebox(0,0)[lb]{\smash{\small 2}}}
\put(100,570){\makebox(0,0)[lb]{\smash{\small 3}}}
\put(120,570){\makebox(0,0)[lb]{\smash{\small 4}}}
\put(140,570){\makebox(0,0)[lb]{\smash{\small 5}}}
\put(160,570){\makebox(0,0)[lb]{\smash{\small 6}}}
\put(180,570){\makebox(0,0)[lb]{\smash{\small 7}}}
\put(200,570){\makebox(0,0)[lb]{\smash{\small 8}}}
\put(220,570){\makebox(0,0)[lb]{\smash{\small 8}}}
\put(240,570){\makebox(0,0)[lb]{\smash{\small 10}}}
\put(260,570){\makebox(0,0)[lb]{\smash{\small 11}}}
\put(280,570){\makebox(0,0)[lb]{\smash{\small 12}}}
\put(300,570){\makebox(0,0)[lb]{\smash{\small 13}}}
\put(320,566){\makebox(0,0)[lb]{\smash{\small 14}}}
\put(340,566){\makebox(0,0)[lb]{\smash{\small 15}}}
\put(220,810){\makebox(0,0)[lb]{\smash{\Large {\bf Figure   4.}}}}
\put(105,695){\makebox(0,0)[lb]{\smash{\bf (k}}}
\put(130,695){\makebox(0,0)[lb]{\smash{\bf , l}}}
\put(155,695){\makebox(0,0)[lb]{\smash{\bf )}}}
\put( 60,720){\makebox(0,0)[lb]{\smash{\bf (k}}}
\put( 75,715){\makebox(0,0)[lb]{\smash{\bf 1}}}
\put( 85,720){\makebox(0,0)[lb]{\smash{\bf , l}}}
\put(100,715){\makebox(0,0)[lb]{\smash{\bf 1}}}
\put(110,720){\makebox(0,0)[lb]{\smash{\bf )}}}
\put(150,675){\makebox(0,0)[lb]{\smash{\bf (k}}}
\put(175,675){\makebox(0,0)[lb]{\smash{\bf , l}}}
\put(200,675){\makebox(0,0)[lb]{\smash{\bf )}}}
\put(120,690){\makebox(0,0)[lb]{\smash{\bf 2}}}
\put(145,690){\makebox(0,0)[lb]{\smash{\bf 2}}}
\put(165,670){\makebox(0,0)[lb]{\smash{\bf 3}}}
\put(190,670){\makebox(0,0)[lb]{\smash{\bf 3}}}
\end{picture}

\pagebreak
\setlength{\unitlength}{0.012500in}%
\begin{picture}(350,268)(5,515)
\thicklines
\put( 77,737){$\bullet$}
\put(137,697){$\bullet$}
\put(197,657){$\bullet$}
\put(257,617){$\bullet$}
\put(320,580){\circle{5}}
\put( 40,565){\vector( 0, 1){265}}
\put( 40,760){\line( 1, 0){  5}}
\put( 20,580){\vector( 1, 0){335}}
\put( 40,820){\line( 1, 0){  5}}
\put( 40,700){\line( 1, 0){  5}}
\put( 40,640){\line( 1, 0){  5}}
\put(100,585){\line( 0,-1){  5}}
\put(100,580){\line( 0, 1){  5}}
\put( 40,600){\line( 1, 0){  5}}
\put( 40,620){\line( 1, 0){  5}}
\put( 40,660){\line( 1, 0){  5}}
\put( 40,680){\line( 1, 0){  5}}
\put( 40,720){\line( 1, 0){  5}}
\put( 40,740){\line( 1, 0){  5}}
\put( 40,780){\line( 1, 0){  5}}
\put( 40,800){\line( 1, 0){  5}}
\put(140,585){\line( 0,-1){  5}}
\put(140,580){\line( 0, 1){  5}}
\put( 60,585){\line( 0,-1){  5}}
\put( 60,580){\line( 0, 1){  5}}
\put( 80,585){\line( 0,-1){  5}}
\put( 80,580){\line( 0, 1){  5}}
\put(120,585){\line( 0,-1){  5}}
\put(120,580){\line( 0, 1){  5}}
\put(180,585){\line( 0,-1){  5}}
\put(180,580){\line( 0, 1){  5}}
\put(160,585){\line( 0,-1){  5}}
\put(160,580){\line( 0, 1){  5}}
\put(180,585){\line( 0,-1){  5}}
\put(180,580){\line( 0, 1){  5}}
\put(180,585){\line( 0,-1){  5}}
\put(180,580){\line( 0, 1){  5}}
\put(200,585){\line( 0,-1){  5}}
\put(200,580){\line( 0, 1){  5}}
\put(220,585){\line( 0,-1){  5}}
\put(220,580){\line( 0, 1){  5}}
\put(240,585){\line( 0,-1){  5}}
\put(240,580){\line( 0, 1){  5}}
\put(260,585){\line( 0,-1){  5}}
\put(260,580){\line( 0, 1){  5}}
\put(280,585){\line( 0,-1){  5}}
\put(280,580){\line( 0, 1){  5}}
\put(300,585){\line( 0,-1){  5}}
\put(300,580){\line( 0, 1){  5}}
\put(320,585){\line( 0,-1){  5}}
\put(320,580){\line( 0, 1){  5}}
\put(340,585){\line( 0,-1){  5}}
\put(340,580){\line( 0, 1){  5}}
\put( 20,580){\line(-1, 0){ 15}}
\put( 20,585){\line( 0,-1){  5}}
\put( 20,580){\line( 0, 1){  5}}
\put(320,580){\line( 0, 1){  0}}
\put(335,570){\line(-3, 2){315}}
\put(240,800){\line( 1, 0){ 65}}
\put(305,800){\line( 1, 0){  5}}
\put( 30,600){\makebox(0,0)[lb]{\smash{\small 1}}}
\put( 30,620){\makebox(0,0)[lb]{\smash{\small 2}}}
\put( 30,640){\makebox(0,0)[lb]{\smash{\small 3}}}
\put( 30,660){\makebox(0,0)[lb]{\smash{\small 4}}}
\put( 30,680){\makebox(0,0)[lb]{\smash{\small 5}}}
\put( 30,700){\makebox(0,0)[lb]{\smash{\small 6}}}
\put( 30,720){\makebox(0,0)[lb]{\smash{\small 7}}}
\put( 30,740){\makebox(0,0)[lb]{\smash{\small 8}}}
\put( 30,760){\makebox(0,0)[lb]{\smash{\small 9}}}
\put( 60,570){\makebox(0,0)[lb]{\smash{\small 1}}}
\put( 15,570){\makebox(0,0)[lb]{\smash{\small -1}}}
\put( 80,570){\makebox(0,0)[lb]{\smash{\small 2}}}
\put(100,570){\makebox(0,0)[lb]{\smash{\small 3}}}
\put(120,570){\makebox(0,0)[lb]{\smash{\small 4}}}
\put(140,570){\makebox(0,0)[lb]{\smash{\small 5}}}
\put(160,570){\makebox(0,0)[lb]{\smash{\small 6}}}
\put(180,570){\makebox(0,0)[lb]{\smash{\small 7}}}
\put(200,570){\makebox(0,0)[lb]{\smash{\small 8}}}
\put(220,570){\makebox(0,0)[lb]{\smash{\small 8}}}
\put(240,570){\makebox(0,0)[lb]{\smash{\small 10}}}
\put(260,570){\makebox(0,0)[lb]{\smash{\small 11}}}
\put(280,570){\makebox(0,0)[lb]{\smash{\small 12}}}
\put(300,570){\makebox(0,0)[lb]{\smash{\small 13}}}
\put( 25,820){\makebox(0,0)[lb]{\smash{\small 12}}}
\put( 25,800){\makebox(0,0)[lb]{\smash{\small 11}}}
\put( 25,780){\makebox(0,0)[lb]{\smash{\small 10}}}
\put(320,565){\makebox(0,0)[lb]{\smash{\small 14}}}
\put(340,565){\makebox(0,0)[lb]{\smash{\small 15}}}
\put( 65,760){\makebox(0,0)[lb]{\smash{\bf (k}}}
\put( 80,755){\makebox(0,0)[lb]{\smash{\bf 1}}}
\put( 90,760){\makebox(0,0)[lb]{\smash{\bf , l}}}
\put(105,755){\makebox(0,0)[lb]{\smash{\bf 1}}}
\put(115,760){\makebox(0,0)[lb]{\smash{\bf )}}}
\put(125,720){\makebox(0,0)[lb]{\smash{\bf (k}}}
\put(150,720){\makebox(0,0)[lb]{\smash{\bf , l}}}
\put(175,720){\makebox(0,0)[lb]{\smash{\bf )}}}
\put(185,680){\makebox(0,0)[lb]{\smash{\bf (k}}}
\put(210,680){\makebox(0,0)[lb]{\smash{\bf , l}}}
\put(235,680){\makebox(0,0)[lb]{\smash{\bf )}}}
\put(250,635){\makebox(0,0)[lb]{\smash{\bf (k}}}
\put(275,635){\makebox(0,0)[lb]{\smash{\bf , l}}}
\put(300,635){\makebox(0,0)[lb]{\smash{\bf )}}}
\put(265,630){\makebox(0,0)[lb]{\smash{\bf 2}}}
\put(290,630){\makebox(0,0)[lb]{\smash{\bf 2}}}
\put(140,715){\makebox(0,0)[lb]{\smash{\bf 3}}}
\put(165,715){\makebox(0,0)[lb]{\smash{\bf 3}}}
\put(200,675){\makebox(0,0)[lb]{\smash{\bf 4}}}
\put(225,675){\makebox(0,0)[lb]{\smash{\bf 4}}}
\put(240,810){\makebox(0,0)[lb]{\smash{\Large {\bf Figure   5.}}}}
\end{picture}

\setlength{\unitlength}{0.012500in}%
\begin{picture}(350,274)(5,550)
\thicklines
\put( 77,717){$\bullet$}
\put(137,677){$\bullet$}
\put(197,637){$\bullet$}
\put(257,597){$\bullet$}
\put( 77,675){$\bullet$}
\put(137,635){$\bullet$}
\put(197,595){$\bullet$}
\put(260,560){\circle{5}}
\put( 40,565){\vector( 0, 1){265}}
\put( 40,760){\line( 1, 0){  5}}
\put( 20,580){\vector( 1, 0){335}}
\put( 40,820){\line( 1, 0){  5}}
\put( 40,700){\line( 1, 0){  5}}
\put( 40,640){\line( 1, 0){  5}}
\put(100,585){\line( 0,-1){  5}}
\put(100,580){\line( 0, 1){  5}}
\put( 40,600){\line( 1, 0){  5}}
\put( 40,620){\line( 1, 0){  5}}
\put( 40,660){\line( 1, 0){  5}}
\put( 40,680){\line( 1, 0){  5}}
\put( 40,720){\line( 1, 0){  5}}
\put( 40,740){\line( 1, 0){  5}}
\put( 40,780){\line( 1, 0){  5}}
\put( 40,800){\line( 1, 0){  5}}
\put(140,585){\line( 0,-1){  5}}
\put(140,580){\line( 0, 1){  5}}
\put( 60,585){\line( 0,-1){  5}}
\put( 60,580){\line( 0, 1){  5}}
\put( 80,585){\line( 0,-1){  5}}
\put( 80,580){\line( 0, 1){  5}}
\put(120,585){\line( 0,-1){  5}}
\put(120,580){\line( 0, 1){  5}}
\put(180,585){\line( 0,-1){  5}}
\put(180,580){\line( 0, 1){  5}}
\put(160,585){\line( 0,-1){  5}}
\put(160,580){\line( 0, 1){  5}}
\put(180,585){\line( 0,-1){  5}}
\put(180,580){\line( 0, 1){  5}}
\put(180,585){\line( 0,-1){  5}}
\put(180,580){\line( 0, 1){  5}}
\put(200,585){\line( 0,-1){  5}}
\put(200,580){\line( 0, 1){  5}}
\put(220,585){\line( 0,-1){  5}}
\put(220,580){\line( 0, 1){  5}}
\put(240,585){\line( 0,-1){  5}}
\put(240,580){\line( 0, 1){  5}}
\put(260,585){\line( 0,-1){  5}}
\put(260,580){\line( 0, 1){  5}}
\put(280,585){\line( 0,-1){  5}}
\put(280,580){\line( 0, 1){  5}}
\put(300,585){\line( 0,-1){  5}}
\put(300,580){\line( 0, 1){  5}}
\put(320,585){\line( 0,-1){  5}}
\put(320,580){\line( 0, 1){  5}}
\put(340,585){\line( 0,-1){  5}}
\put(340,580){\line( 0, 1){  5}}
\put( 20,580){\line(-1, 0){ 15}}
\put( 20,585){\line( 0,-1){  5}}
\put( 20,580){\line( 0, 1){  5}}
\put(320,560){\line(-3, 2){300}}
\put(260,800){\line( 1, 0){ 75}}
\multiput(259,559)(-25.33128,16.88752){10}{\line(-3, 2){ 12.019}}
\put( 30,600){\makebox(0,0)[lb]{\smash{\small 1}}}
\put( 30,620){\makebox(0,0)[lb]{\smash{\small 2}}}
\put( 30,640){\makebox(0,0)[lb]{\smash{\small 3}}}
\put( 30,660){\makebox(0,0)[lb]{\smash{\small 4}}}
\put( 30,680){\makebox(0,0)[lb]{\smash{\small 5}}}
\put( 30,700){\makebox(0,0)[lb]{\smash{\small 6}}}
\put( 30,720){\makebox(0,0)[lb]{\smash{\small 7}}}
\put( 30,740){\makebox(0,0)[lb]{\smash{\small 8}}}
\put( 30,760){\makebox(0,0)[lb]{\smash{\small 9}}}
\put( 26,780){\makebox(0,0)[lb]{\smash{\small 10}}}
\put( 26,800){\makebox(0,0)[lb]{\smash{\small 11}}}
\put( 26,820){\makebox(0,0)[lb]{\smash{\small 12}}}
\put( 60,570){\makebox(0,0)[lb]{\smash{\small 1}}}
\put( 15,570){\makebox(0,0)[lb]{\smash{\small -1}}}
\put( 80,570){\makebox(0,0)[lb]{\smash{\small 2}}}
\put(100,570){\makebox(0,0)[lb]{\smash{\small 3}}}
\put(120,570){\makebox(0,0)[lb]{\smash{\small 4}}}
\put(140,570){\makebox(0,0)[lb]{\smash{\small 5}}}
\put(160,570){\makebox(0,0)[lb]{\smash{\small 6}}}
\put(180,570){\makebox(0,0)[lb]{\smash{\small 7}}}
\put(200,570){\makebox(0,0)[lb]{\smash{\small 8}}}
\put(220,570){\makebox(0,0)[lb]{\smash{\small 8}}}
\put(280,570){\makebox(0,0)[lb]{\smash{\small 12}}}
\put(320,570){\makebox(0,0)[lb]{\smash{\small 14}}}
\put(340,570){\makebox(0,0)[lb]{\smash{\small 15}}}
\put(260,810){\makebox(0,0)[lb]{\smash{\Large {\bf Figure   6.}}}}
\put(260,570){\makebox(0,0)[lb]{\smash{\small 11}}}
\put( 60,655){\makebox(0,0)[lb]{\smash{\small {\bf 2}}}}
\put( 66,660){\makebox(0,0)[lb]{\smash{\small {\bf ,l'}}}}
\put( 47,660){\makebox(0,0)[lb]{\smash{\small {\bf (k'}}}}
\put( 75,655){\makebox(0,0)[lb]{\smash{\small {\bf 2}}}}
\put( 81,660){\makebox(0,0)[lb]{\smash{\small {\bf )}}}}
\put(109,635){\makebox(0,0)[lb]{\smash{\small {\bf ,l'}}}}
\put( 90,635){\makebox(0,0)[lb]{\smash{\small {\bf (k'}}}}
\put(125,635){\makebox(0,0)[lb]{\smash{\small {\bf )}}}}
\put(103,630){\makebox(0,0)[lb]{\smash{\small {\bf 3}}}}
\put(119,630){\makebox(0,0)[lb]{\smash{\small {\bf 3}}}}
\put(169,595){\makebox(0,0)[lb]{\smash{\small {\bf ,l'}}}}
\put(150,595){\makebox(0,0)[lb]{\smash{\small {\bf (k'}}}}
\put(185,595){\makebox(0,0)[lb]{\smash{\small {\bf )}}}}
\put(163,590){\makebox(0,0)[lb]{\smash{\small {\bf 1}}}}
\put(179,590){\makebox(0,0)[lb]{\smash{\small {\bf 1}}}}
\put(125,700){\makebox(0,0)[lb]{\smash{\bf (k}}}
\put(150,700){\makebox(0,0)[lb]{\smash{\bf , l}}}
\put(175,700){\makebox(0,0)[lb]{\smash{\bf )}}}
\put( 65,740){\makebox(0,0)[lb]{\smash{\bf (k}}}
\put( 80,735){\makebox(0,0)[lb]{\smash{\bf 2}}}
\put( 90,740){\makebox(0,0)[lb]{\smash{\bf , l}}}
\put(105,735){\makebox(0,0)[lb]{\smash{\bf 2}}}
\put(115,740){\makebox(0,0)[lb]{\smash{\bf )}}}
\put(185,655){\makebox(0,0)[lb]{\smash{\bf (k}}}
\put(210,655){\makebox(0,0)[lb]{\smash{\bf , l}}}
\put(235,655){\makebox(0,0)[lb]{\smash{\bf )}}}
\put(255,615){\makebox(0,0)[lb]{\smash{\bf (k}}}
\put(280,615){\makebox(0,0)[lb]{\smash{\bf , l}}}
\put(305,615){\makebox(0,0)[lb]{\smash{\bf )}}}
\put(270,610){\makebox(0,0)[lb]{\smash{\bf 1}}}
\put(295,610){\makebox(0,0)[lb]{\smash{\bf 1}}}
\put(200,650){\makebox(0,0)[lb]{\smash{\bf 3}}}
\put(225,650){\makebox(0,0)[lb]{\smash{\bf 3}}}
\put(140,695){\makebox(0,0)[lb]{\smash{\bf 4}}}
\put(165,695){\makebox(0,0)[lb]{\smash{\bf 4}}}
\end{picture}
\pagebreak

\end{document}